\documentstyle[prb,aps,multicol,epsf]{revtex}
\tighten
\begin{document}
\draft

\title{Self-consistent theory of shot noise in nondegenerate ballistic
conductors}
\author{O. M. Bulashenko and J. M. Rub\'{\i}}
\address{Departament de F\'{\i}sica Fonamental,
Universitat de Barcelona, Diagonal 647, E-08028 Barcelona, Spain}
\author{V. A. Kochelap}
\address{Department of Theoretical Physics,
Institute of Semiconductor Physics, Kiev 252028, Ukraine}
\date{\today}
\maketitle

\begin{abstract}
A self-consistent theory of shot noise in ballistic two-terminal
conductors under the action of long-range Coulomb correlations is presented.
Analytical formulas for the electron distribution function and its fluctuation
along the conductor, which account for the Coulomb correlations,
have been derived.
Based upon these formulas, the current-noise reduction factor has been 
obtained for biases ranging from thermal to shot-noise limits as dependent
on two parameters: the ratio between the length of the sample and
the Debye screening length $\lambda=d/L_D$ and the applied voltage $qU/k_BT$.
The difference with the formulas for a vacuum diode is discussed.
\end{abstract}


\begin{multicols}{2}

\section{Introduction}

Recently, significant attention has been focused on the study of
nonequilibrium fluctuations of current (shot noise) in mesoscopic
conductors. \cite{dejong97}
The term ``shot noise,'' appearing originally in the context of pure
ballistic electron transmission in vacuum-tube devices, \cite{schottky18}
has acquired nowadays a much broader usage and refers to different mesoscopic
structures, including diffusive conductors and resonant-tunneling devices,
where the carrier flow exhibits nonequilibrium noise proportional to
the electric current. \cite{dejong97}

A matter of particular interest is the significance of long-range Coulomb
correlations in the noise-reduction effect. \cite{landauer,buttiker}
Coulomb interactions may keep nearby electrons apart and more regularly spaced
rather than strictly at random, which leads to the noise reduction, 
as pointed out by Landauer. \cite{landauer}
This effect occurs in different physical situations.
Among them are charge-limited ballistic transport, resonant tunneling,
single-electron tunneling, etc.
For the ballistic conductors an electrostatic potential barrier is formed
near an injecting contact. The barrier fluctuates synchronously with
random electron passages through it, which leads to noise reduction, as
evidenced recently by Monte Carlo simulations for semiconductor ballistic
diodes. \cite{gonzalez97} In this way, an incoming Poissonian flow is
converted into an outgoing sub-Poissonian flow, exhibiting a motional
electron-number squeezing. \cite{b98}
This effect is similar to that leading to shot-noise suppression
in vacuum diodes. \cite{rack38,north40,ziel54}
Under the resonant tunneling effect, a built-in charge inside a quantum
well affects the position of the resonant level and prevents the incoming
carriers from passing through the well, thereby resulting in carrier
correlation and shot-noise reduction \cite{li90b,brown92,liu95hc}
in a certain range of biases. \cite{iannacone98,blanter98}
The Coulomb correlations in these systems act under the coherent as well as
under the sequential tunneling regime of the carrier transport.
The carrier correlations reach their extreme form of the Coulomb blockade of
the electron transfer under the single-electron tunneling effect,
leading to the noise reduction studied theoretically
\cite{korotkov92,hershfield93,hanke93,imamoglu93,yamanishi96,matsuoka98}
and observed in experiment. \cite{birk95}

All the above-mentioned cases have the common features which are necessary
for the Coulomb regulation effect and shot-noise reduction {\em in the whole
frequency spectrum} to occur:
(i) the existence of a potential barrier inside a device or at the interface
with an injecting electron reservoir, which controls the current;
(ii) the dependence of the barrier height and/or carrier transmission on
the current.
If no barrier is present, no shot-noise reduction at low frequencies
due to Coulomb repulsion is expected.
At high frequencies, however, the noise level may also be affected by
Coulomb correlations due to screening in an external environment.
\cite{naveh97,nagaev98}

The potential barrier, which controls the current, appears in an
ordinary situation of the space-charge-limited transport.
For {\em ballistic} nondegenerate conductors this case has been treated 
recently by Monte Carlo simulations \cite{gonzalez97,b98} and attracted 
some attention in Ref.\ \onlinecite{naveh99} for a degenerate case.
For the case of {\em diffusive} nondegenerate conductors,
studied by the Monte Carlo technique in Ref.\ \onlinecite{gonzalez98a},
the self-consistent kinetic theory of noise,
which takes into account Coulomb correlations, has been developed recently 
in Refs.\ \onlinecite{beenakker99} and \onlinecite{schomerus99}.
A similar kinetic theory for the ballistic case is lacking.

It is the aim of this paper to address the problem
of Coulomb correlations in {\em ballistic} conductors and present
for the first time a self-consistent theory of shot noise
in these conductors by solving analytically the kinetic equation coupled
self-consistently with the Poisson equation.
It is important to compare the present noise theory for a semiconductor
ballistic diode with that for a vacuum diode developed long ago.
\cite{north40,ziel54}
The main advance for the latter has been done in the celebrated paper
by North published in 1940, where he derived an asymptotic formula for
the current-noise spectral density at the high voltage limit. \cite{north40}
Monte Carlo simulations of noise in vacuum diodes are also available.
\cite{tien56,wen64,saito72}
It should be stressed, however, that despite of the similarity of
the underlying physics (in both cases the nondegenerate Boltzmann electron
gas without collisions in the electrostatic field is under consideration),
the case of the semiconductor diode differs by several features:
(i) due to a two-terminal geometry,
there are two opposing currents instead of a single current, which results
in different current-voltage characteristics at low and moderate biases;
\cite{remark1}
(ii) the ballistic transport regime is limited by the presence of disorder,
impurities, etc.
Even in a pure and perfect solid, carriers may interact with a lattice
(phonons), which at high biases becomes significant and breaks down
the ballistic regime.
This makes it practically impossible to attain in solids the regime where
the known formulas for vacuum electronics, such as the Child law for $I$-$V$
characteristics or North's asymptotic formula for the noise,
may be applied.
This issue will be addressed in the paper, using the derived formulas
and considering them in a full range of biases.
Finally, we suggest an electron spectroscopy experiment to make the Coulomb 
correlations effect observable. The possibility of such an experiment is 
based on recent advances in nanoscale fabrication techniques and shot noise 
measurements. \cite{reznikov95,kumar96,schoel97}

The paper is organized as follows.
In Sec.\ II we describe the semiconductor ballistic structure
and discuss the main assumptions concerning underlying physics.
In particular, the validity of the one-dimensional plane geometry
approximation for the fluctuation problem is addressed.
In Sec.\ III we introduce the basic equations that describe
the space-charge-limited semiclassical transport: the collisionless kinetic
equation coupled self-consistently with the Poisson equation.
The steady-state problem is solved in Sec.\ IV, and the results are compared
with the Monte Carlo simulations.
In Sec.\ V we solve analytically the fluctuation problem and derive
the formula for the current-noise spectral density that covers the range of
biases from thermal to the shot-noise limits.
The results for the noise-reduction factor are compared with Monte Carlo
simulations and North's asymptotic formula for vacuum diodes.
The contributions of different electron energy groups to the noise are found,
and the correlations in energies for the electrons collected at the receiving
contact are discussed.
Finally, Sec.\ VI summarizes the main contributions of the paper,
and in the Appendix we present mathematical details concerning
the derivation of the fluctuations of the electron distribution function
in the self-consistent electric field.

\section{The physical model}

\begin{figure}
\narrowtext
\epsfxsize=8.0cm
\epsfbox{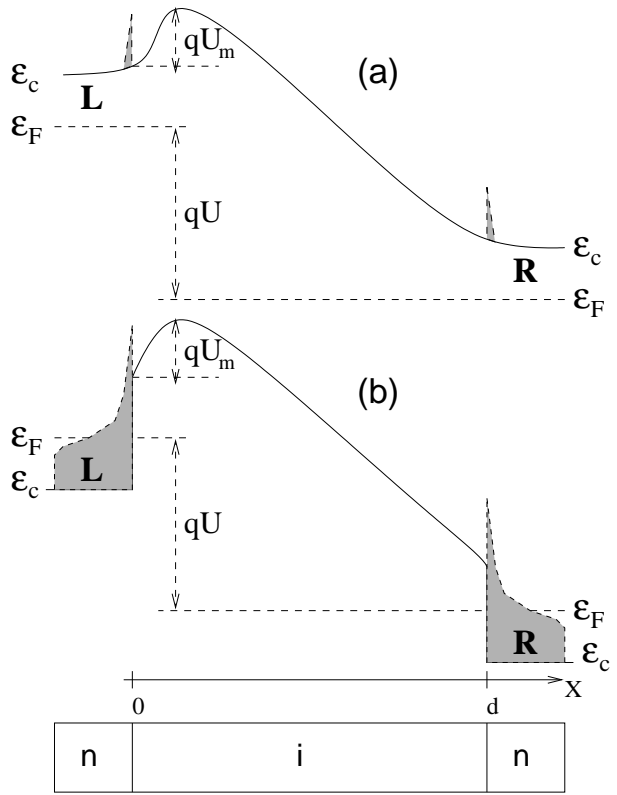}
\protect\vspace{0.5cm}
\caption{
Schematic band-energy diagram for a $n$-$i$-$n$ ballistic diode under
a space-charge-limited conduction.
Two different types of the contacts are shown:
(a) homojunctions; (b) heterojunctions.
Shadowed regions illustrate the energy distribution function
of electrons at the contact-sample interfaces.}
\label{f1} \end{figure}

Before proceeding with a discussion of the problem, we will specify
the structure under consideration and the main assumptions concerning
the underlying physics.
Consider a two-terminal semiconductor ballistic sample with plane parallel
contacts at $X=0$ and $X=d$ (see Fig.~\ref{f1}).
The contacts, which we denote by $L$ and $R$ (left and right),
are assumed to be heavily doped semiconductors with a contact electron
density much higher than that in the sample.

\noindent
The structure may then be considered as a $n$-$i$-$n$ diode operating under
a space-charge-limited current regime in which the current is determined by
a charge injection from the contacts rather than by intrinsic carriers of
the active region. \cite{lampert70}
Two different types of the contacts may be considered depending on whether
the contact and the sample are fabricated of the same or different material.
For the former case the diode is composed of two homojunctions
[Fig.\ \ref{f1}(a)], while for the latter, it is composed of two
heterojunctions with a jump of the conductance band $\varepsilon_c$
at the contact-sample interface [Fig.\ \ref{f1}(b)].
The underlying physics is similar if in both cases the contact doping
is such that the Fermi level $\varepsilon_F$ is sufficiently below
the edge of the conduction band in the sample.
In such a case, only the tail of the distribution function is injected,
which leads to the nondegeneracy of the electron gas in the ballistic part
of the diode.
The theory is therefore applicable to quantum heterostructures with
over-barrier transport, \cite{mitin99} where current is determined by a tail
in the distribution function
(ballistic-injection, real-space-transfer devices, etc.), as well as
for the homodiode with a nondegenerate electron gas in the contacts.

In order to simplify the problem, we assume that under the range of biases
of interest, due to the large difference in the carrier density between
the contacts and the sample, and hence in the corresponding Debye screening
lengths, all the band bending occurs in the ballistic base,
and therefore the relative position of the conduction band and the Fermi
level $\varepsilon_c-\varepsilon_F$ does not change in the contacts.
For such a modeling, all of the potential drop takes place exclusively inside
the ballistic base between the positions $X=0$ and $X=d$ in Fig.\ \ref{f1},
and the contacts may be excluded from the consideration.
This assumption is better fulfilled for the case of the heterojunctions
because of much higher electron densities in the contacts.

The carriers inside the contacts are assumed to remain at thermal equilibrium,
and their injected part is distributed over the energy according to
the Maxwell-Boltzmann distribution function at lattice temperature $T$.
For the ballistic part of the diode, we suppose

\begin{equation}
\lambda_w\ll d\lesssim \lambda_p,
\end{equation}
with $\lambda_w$ the electron
wavelength and $\lambda_p$ the mean free path, so that electrons may be
considered as classical particles moving ballistically between the contacts
and interacting with each other electrostatically.
This regime is accessible in modern device fabrication technologies
for which the mean free path $\lambda_p$ may be as high as
$10^4$--$10^5\,{\rm nm}$ in modulation-doped structures (for instance,
in GaAs/Al$_x$Ga$_{1-x}$As at low temperatures \cite{beenakker91,facer99})
and $\sim 10^3\,{\rm nm}$ in the purest bulk material,
whereas the Fermi wavelength is about $40\,{\rm nm}$.

Next we assume that the transversal size of the diode is sufficiently thick
(much larger than the screening length $L_D$).
This allows us to treat the steady-state electrostatic problem as a
one-dimensional one in the plane geometry.
However, to use the same one-dimensional consideration for the fluctuation
problem, we need an additional justification.
The fluctuating current is determined by a random transmission
of discrete electron charges of the amount of $q$.
Essentially, this discreteness of charge transmission together with
randomness leads to the shot noise.
In principle, each single electron while transmitted between the contacts
disturbs the electric field and thereby interacts with other electrons
of the current flow in both longitudinal and transversal directions.
The electrostatic screening in such a problem is three-dimensional.
Nevertheless, we shall treat the problem as a
one-dimensional one considered in the plane geometry
by averaging the fluctuations over the transversal directions.
This is justified if the average distance between the excess (fluctuating)
carriers in transversal direction is much smaller than the characteristic 
scale of the electrostatic potential variation in that direction.
This condition may be written as

\begin{equation} \label{condpl}
L_{\perp}^2 d \sqrt{\langle\delta n^2\rangle} \gg 1,
\end{equation}
where $L_{\perp}$ is the transverse characteristic scale,
$n$ is the typical electron density in the ballistic region, and
$\delta n$ its fluctuation.
To estimate the order of magnitude of the fluctuation $\delta n$,
we use Poissonian statistics, leading to the relation
$\langle\delta n^2\rangle\sim n/(L_{\perp}^2 d)$.
Thus, the condition (\ref{condpl}) becomes $n L_{\perp}^2 d \gg 1$.
The scale $L_{\perp}$ depends on the ratio between the longitudinal dimension
$d$ of the sample and the Debye screening length $L_D$ in the active region.
For nondegenerate electrons, the latter is defined
as $L_D=\sqrt{\kappa k_B T/(q^2 n)}$, with $\kappa$ being the dielectric
permittivity and $k_B$ the Boltzmann constant.
To estimate the magnitude of $L_{\perp}$, we distinguish two different cases: 
(i) Weak screening, $d \lesssim L_D$: 
For this case $L_{\perp}\sim d$, \cite{remark2}
and the condition (\ref{condpl}) becomes $n\gg d^{-3}$, which
for $d\sim 300$ nm requires $n\gg 10^{14}\,{\rm cm}^{-3}$.
(ii) Strong screening, $d\gg L_D$: 
For this case $L_{\perp}\sim L_D$,
and condition (\ref{condpl}) becomes $n\gg L_D^{-2} d^{-1}$.
After the substitution of the expression for the screening length,
it is seen that this condition becomes independent of $n$, although
it requires a sufficiently long sample,
$d\gg q^2/(\kappa k_BT) \sim 2 a_0 (E_0/k_BT)$.
where $a_0=\kappa\hbar^2/(mq^2)$ is the effective Bohr radius and
$E_0=q^2/(2\kappa a_0)$ is the effective Rydberg energy in the material.
For GaAs, $a_0\approx 10$ nm, $E_0\approx 5$ meV,
which corresponds to the temperature of about $60\,{\rm K}$.
Then for $T\sim 10$ K, $d\gg 120$ nm, which is supposed to be fulfilled.
On another hand, the condition of strong screening requires $d\gg L_D$,
which leads to the condition on the electron density

\begin{equation} \label{condpl2}
n\gg \frac{k_BT}{E_0} \frac{1}{2a_0 d^2}.
\end{equation}
For the same set of parameters, one gets $n\gg 2\times 10^{14}\,{\rm cm}^{-3}$.
Therefore, for both cases of weak and strong screening, there is
a requirement on the minimal electron density or, equivalently, on
the minimal density of the injection current in order to use
the one-dimensional electrostatic screening picture for the fluctuations.
Otherwise, each carrier perturbs the electrostatic potential
independently and the three-dimensional approach is needed.
On the other hand, the assumption of the nondegenerate electron gas restricts
our approach by a maximum electron concentration dependent on $T$.
For temperatures in the range 10--77 K, these maximal concentrations
are estimated to be in the range $3\times 10^{16}$ to
$6\times 10^{17}\,{\rm cm}^{-3}$.
These estimates show that the approach undertaken below covers a wide
range of typical diode parameters: electron concentrations,
diode lengths, and temperatures.


\section{Basic equations}
A semiclassical space-charge-limited transport in a ballistic conductor is
completely described by the electron distribution function
$F(X,v_x,t)$ and the electrostatic potential $\varphi(X,t)$.
Here, $v_x$ is the $X$ component of the electron velocity and $t$ is the time.
The potential $\varphi(X,t)$ inside the sample is determined
by the distribution of space charge from the Poisson equation

\begin{equation} \label{pois}
\frac{d^2\varphi}{dX^2} = \frac{q}{\kappa} N(X,t),
\end{equation}
with the boundary conditions

\begin{equation} \label{bcpois}
\varphi(0,t)=\varphi_L, \quad \varphi(d,t)=\varphi_R.
\end{equation}
The voltage bias between the contacts
$U=\varphi_R-\varphi_L$ is assumed to be fixed by
a low-impedance external circuit.
The electron density $N(X,t)$ at any plane $X$ is determined by integrating
the local electron distribution function over velocities

\begin{equation} \label{nxgen}
N(X,t) = \int_{-\infty}^{\infty} F(X,v_x,t) dv_x,
\end{equation}
whereas the current in the external lead is given by \cite{ziel54}
\begin{equation} \label{igen}
I(t) = -\frac{q A}{d} \int_0^d
\left[ \int_{-\infty}^{\infty} v_x F(X,v_x,t) dv_x \right] dX
+ C_0 \frac{\partial U}{\partial t},
\end{equation}
where $C_0=\kappa A/d$ is a capacitance and $A$ the cross-sectional area.
Due to a fixed-applied-voltage condition, in what follows we shall neglect
the last term in Eq.\ (\ref{igen}) coming from the displacement current
contribution. In addition, for simplicity, we shall omit the minus sign
for the current, which is opposed to the direction of electron flow.
Moreover, as will be shown below, the current is conserved along the sample
due to the conservation of electron energy under ballistic motion
(this is true for both the stationary current and its fluctuation).
Therefore, the integration over $X$ becomes trivial and it will be disregarded.

Under ballistic motion the distribution function $F(X,v_x,t)$ obeys
the collisionless kinetic equation

\begin{equation} \label{vlas}
\frac{\partial F}{\partial t}
+ v_x \frac{\partial F}{\partial X} +
\frac{q}{m}\frac{d\varphi}{dX} \frac{\partial F}{\partial v_x} = 0,
\end{equation}
where $m$ stands for the electron effective mass.
The distribution functions of injected from the contacts electrons
are assumed to be given as

\begin{eqnarray} \label{bcvlas}
F(0,v_x,t)|_{v_x>0} &=& F_L(v_x,t), \nonumber\\
F(d,v_x,t)|_{v_x<0} &=& F_R(v_x,t).
\end{eqnarray}
The kinetic equation (\ref{vlas}) with the electrostatic potential
determined self-consistently from Eqs.\ (\ref{pois}) and (\ref{nxgen})
are known as the Vlasov system of equations \cite{vlasov38} describing
the dynamical screening of the interaction in plasma. \cite{balescu75}

Equation (\ref{vlas}) may also be expressed as

\begin{equation}
\left.\frac{d F}{d t}\right|_{trajectory} = 0,
\end{equation}
since $F$ is constant along an electron trajectory, i.e., the
distribution function at any plane $X$ can be expressed through
the functions $F_k(v_x,t)$, $k=L,R$ defined at the boundaries.
Each of these functions is considered to consist of two terms,
a stationary part describing the stationary injection and
a time-varying stochastic component.
Explicitly,

\begin{equation} \label{bcdft}
F_k(v_x,t) = \bar{F_k} (v_x) + \delta F_k(v_x,t),
\quad k=L,R.
\end{equation}
Under nondegenerate and equilibrium conditions in the contacts,
we assume for the stationary part of the injection function
the half-Maxwellian distribution

\begin{equation} \label{max0}
\bar{F_k}(v_x) = \frac{2 N_0}{v_0\sqrt{\pi}} e^{-v_x^2/v_0^2}
\end{equation}
with $v_x > 0$ for $k=L$ and $v_x < 0$ for $k=R$.
Here, $N_0$ is the density of electrons injected from the contacts and
$v_0=\sqrt{2k_B T/m}$ is the thermal velocity.
The contact distribution functions (\ref{max0}) are normalized in such 
a way that the integration over a half-velocity space yields the density 
of electrons injected from the contact

\begin{equation}
N_0=\int_{v_x > 0} \bar{F_L}(v_x) d v_x=\int_{v_x < 0} \bar{F_R}(v_x) d v_x.
\end{equation}

The stochastic terms $\delta F_k$, $k=L,R$ in Eq.\ (\ref{bcdft}) are
the only sources of noise under the ballistic transport considered here, 
since the electron motion between the contacts is noiseless.
Their equal-time correlation, due to equilibrium conditions, 
is given by \cite{kogan}

\begin{eqnarray} \label{corrf}
\langle\delta F_k&&(v_x,t)\delta F_{k'}(v_x',t)\rangle \nonumber\\
&&= C \bar{F}(v_x) [1 - \bar{F}(v_x)]
\delta_{kk'}\delta(v_x-v_x'),
\end{eqnarray}
where the constant $C$ is determined from the normalization condition.
Since the injected electron gas is nondegenerate,
$\bar{F}\ll 1$, and the factor $1-\bar{F}$ will be ignored.

As a consequence of the fluctuations inside the contacts
(whose origin is ultimately the carrier scattering processes),
both the electron distribution function and electrostatic potential
in the ballistic sample fluctuate, leading to the current fluctuations.
These quantities will be presented as a sum of stationary and fluctuating
contributions:
$F(X,v_x,t)=\bar{F}(X,v_x)+\delta F(X,v_x,t)$,
$N(X,t)=\bar{N}(X)+\delta N(X,t)$,
$\varphi(X,t)=\bar{\varphi}(X)+\delta\varphi(X,t)$, and
$I(t)=\bar{I}+\delta I(t)$.

Introducing the Fourier transform for the fluctuations of the distribution
function $\delta F_{\omega}(X,v_x)$ and the potential
$\delta\varphi_{\omega}(X)$, the kinetic equation takes on the form

\begin{equation} \label{vlasf}
-i \omega \delta F_{\omega}
+ v_x \frac{\partial \delta F_{\omega}}{\partial X} +
\frac{q}{m} \frac{d\bar{\varphi}}{dX}
\frac{\partial\delta F_{\omega}}{\partial v_x}
+ \frac{q}{m} \frac{\partial \bar{F}}{\partial v_x}
\frac{d\delta\varphi_{\omega}}{dX} = 0,
\end{equation}
with the boundary conditions at the contacts

\begin{eqnarray} \label{bcvlasf}
\delta F_{\omega}(0,v_x)|_{v_x>0} = \delta F_L^{\omega}(v_x), \nonumber\\
\delta F_{\omega}(L,v_x)|_{v_x<0} = \delta F_R^{\omega}(v_x),
\end{eqnarray}
where $\delta F_L^{\omega}$ and $\delta F_R^{\omega}$ are the Fourier
transforms of the stochastic functions from Eq.~(\ref{bcdft}).
The equation for the fluctuating potential $\delta\varphi_{\omega}$
is trivially obtained from Eqs.~(\ref{pois}) and (\ref{nxgen}),

\begin{equation} \label{dpois}
\frac{d^2 \delta\varphi_{\omega}}{dX^2} =
\frac{q}{\epsilon} \int \delta F_{\omega}(X,v_x) d v_x,
\end{equation}
the boundary conditions for which follows from Eq.~(\ref{bcpois}),

\begin{equation} \label{bcdpois}
\delta\varphi_L^{\omega}(0)=0, \quad \delta\varphi_R^{\omega}(d)=0.
\end{equation}

Below we restrict ourselves to the calculation of the low-frequency
plateau of the noise spectrum; thus one can omit the term
proportional to $\omega$ in Eq.~(\ref{vlasf}). It can be shown that this
approximation is valid if the shortest fluctuation period in $\delta F_k(t)$
is considered to be sufficiently greater than the average electron transit
time $\tau_T$ across the diode, i.e., $\omega\ll\tau_T^{-1}$.
Thus, the above self-consistent equations completely describe
the stationary transport and low-frequency fluctuations in the ballistic
sample, and below we shall omit the index $\omega$.

It is advantageous to rescale all the variables as follows:

\begin{eqnarray} \label{norm}
w = \frac{v_x}{v_0}, \quad  x &=& \frac{X}{L_D},
\quad \psi = \frac{q\bar{\varphi}}{k_B T} \nonumber\\
n = \frac{\bar{N}}{2 N_0}, \quad  f &=& \bar{F} \frac{v_0}{2 N_0}, \quad
\delta f = {\delta F} \frac{v_0}{2 N_0}.
\end{eqnarray}
In such units the basic equations contain only two dimensionless parameters:
(i) the length of the sample (or the screening parameter) $\lambda=d/L_D^0$,
where $L_D^0=\sqrt{\epsilon k_B T/(2q^2 N_0)}$ is the Debye screening length
corresponding to the electron density $2N_0$,
and (ii) the applied voltage bias $V=qU/(k_BT)$.
Below we use the dimensionless variables in all the equations.

\section{Steady-state problem}

The calculation of fluctuations in the ballistic conductor requires the
knowledge of the stationary distribution of electrostatic field, which,
in turn, can be determined by solving the full steady-state problem.
The self-consistent steady-state problem can be solved as follows.
First, we solve the stationary collisionless kinetic equation
for the distribution function $f(x,w)$

\begin{equation} \label{vlasns}
w \frac{\partial f}{\partial x} +
\frac{1}{2}\frac{d \psi}{d x} \frac{\partial f}{\partial w} = 0
\end{equation}
at a given electrostatic potential $\psi(x)$.
Integrating $f(x,w)$ over $w$, we then find the electron density profile
$n(\psi)$ in terms of the potential $\psi(x)$.
Then we should solve the Poisson equation

\begin{equation} \label{poisn}
\frac{d^2\psi}{dx^2} = n(\psi),
\end{equation}
with the boundary conditions

\begin{equation} \label{bcpoisn}
\psi(0)\equiv\psi_L=0, \quad \psi(\lambda)\equiv\psi_R=V.
\end{equation}
Here, we set the zero value of the potential at the left contact.

\subsection{Stationary distribution function}
\label{stat-df}

To solve the stationary kinetic equation (\ref{vlasns}),
we have to specify the boundary conditions for this equation at a given
$\psi(x)$. Generally, the {\em nonstationary} kinetic equation (\ref{vlas})
and the distribution functions (\ref{bcvlas}) of injected electrons
completely determine the nonstationary solution $f(x,w,t)$.
However, the {\em steady-state} Equation (\ref{vlasns}) requires 
a specification of the boundary conditions for the distribution function 
of all the electrons: those {\em injected} from the contacts into the sample 
and those {\em leaving} the sample.
Let the space charge in the sample be such that a potential minimum
$\psi_m$ occurs at $x=x_m$, which acts as a potential barrier for electrons.
We define the total electron energy $\varepsilon_t=w^2-\psi(x)$.
For a given potential, the distribution function should consist of the
terms originating from two electron streams injected by the left and right
contacts.
Electrons injected from each of the contacts fall into two groups
depending on their injecting energies.
If the initial energy is higher than the height of the barrier, electrons
obviously reach the opposite contact and contribute to the electric current.
These electrons are not reflected back.
Note that the height of the barrier is different for the electrons
injected from the left and right contacts.
For those injected from the left, it is $\psi_L-\psi_m=V_m$,
which is the potential minimum depth, while for those injected from the right,
it is $\psi_R-\psi_m=V+V_m$ (see Fig.\ \ref{f1}).
Accordingly, the lower bounds for the velocities of the {\em transmitted}
electrons are given by

\begin{eqnarray} \label{wlr}
w_L&=&\sqrt{\psi_L-\psi_m}=\sqrt{V_m}, \nonumber\\
w_R&=&\sqrt{\psi_R-\psi_m}=\sqrt{V_m+V}.
\end{eqnarray}
Electrons of the second group, which we shall call the {\em reflected}
electrons, are reflected by the barrier and do not contribute to
the current (however, both groups affect the electrostatic potential).
An electron from the second group being injected with a velocity $w$
returns to the contact with the opposite velocity  of the same value $-w$.
Taking into account the above consideration, the electron distribution
function $f(x,w)$ at any plane $x$ may be written as

\begin{equation} \label{dfsum}
f = f_{L,t} + f_{L,r} + f_{R,t} + f_{R,r},
\end{equation}
where the indices $L$ and $R$ refer to the left and right contacts, 
and the indices $t$ and $r$ distinguish the transmitted and reflected 
groups of carriers, respectively.
The boundary conditions for these functions read

\begin{eqnarray} \label{bcdf}
f_{L,t}(0,w_c) &=& f_L(w_c) \, \theta(w_c - w_L), \nonumber\\
f_{L,r}(0,w_c) &=& f_L(w_c) \, \theta(w_L^2 - w_c^2), \nonumber\\
f_{R,t}(\lambda,w_c) &=& f_R(w_c) \, \theta(-w_c - w_R), \nonumber\\
f_{R,r}(\lambda,w_c) &=& f_R(w_c) \, \theta(w_R^2 - w_c^2),
\end{eqnarray}
where $w_c$ is the $x$ velocity component of injected electrons at
the contacts, $\theta$ is the Heaviside step function, and
the distribution function of injected electrons is determined by
Eq.~(\ref{max0}), which in dimensionless units reads

\begin{equation} \label{maxn}
f_L(w_c) = f_R(w_c) = \frac{1}{\sqrt{\pi}} e^{-w_c^2}.
\end{equation}

\begin{figure}
\epsfxsize=8.0cm\epsfbox{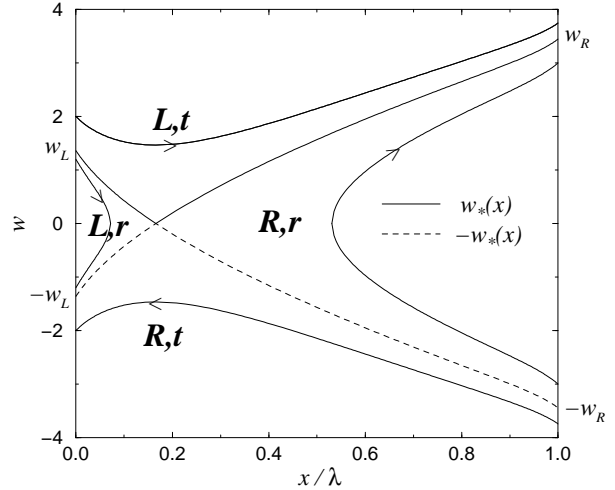}
\protect\vspace{0.5cm}
\narrowtext
\caption{
Typical electron trajectories in the phase space $(x,w)$ for different
electron groups: $L$ and $R$ refer to the carriers originated from the left 
and right contacts, and $t$ and $r$ refer to the transmitted and reflected 
groups of carriers.
The separating curves are the critical velocities $\pm w_*(x)$,
which intersect at the point of the potential minimum $(x_m,0)$.
The results are for $\lambda$=30, $V$=10.
}
\label{wx} \end{figure}

We can solve now the collisionless kinetic equation (\ref{vlasns})
explicitly for a given potential profile $\psi(x)$.
Indeed, one can easily see that its solution is an arbitrary function
dependent on the total electron energy
${\cal F}(\varepsilon_t)={\cal F} \biglb( w^2 - \psi(x) \bigrb)$.
The boundary conditions (\ref{bcdf}) determine the shape of this function.
By using the electron-energy conservation law

\begin{equation} \label{encons}
w^2 - \psi(x)= w_c^2 - \psi_k, \quad k=L,R
\end{equation}
where $w_c$ and $\psi_k$ are the parameters at the contacts,
we exclude $w_c$ in the boundary conditions (\ref{bcdf}) and
obtain the contributions in the distribution function as

\begin{mathletters} \label{df} \begin{eqnarray}
f_{L,t}(x,w) &=& \frac{1}{\sqrt{\pi}} \theta\biglb(w - w_*(x)\bigrb) \,
e^{-w^2+\psi(x)-\psi_L}, \label{df-lt} \\
f_{R,t}(x,w) &=& \frac{1}{\sqrt{\pi}} \theta\biglb(-w - w_*(x)\bigrb) \,
e^{-w^2+\psi(x)-\psi_R}, \label{df-rt} \\
f_{k,r}(x,w) &=& \frac{1}{\sqrt{\pi}} \theta\biglb(w_*^2(x)-w^2\bigrb)\,
e^{-w^2+\psi(x)-\psi_k}, \label{df-r}
\end{eqnarray} \end{mathletters}
where $k=L,R$, and the functions $f_{L,t}$ and $f_{R,t}$ for
the transmitted electrons are defined in the whole range $0<x< \lambda$,
whereas the expressions for the reflected electrons $f_{L,r}$ and $f_{R,r}$
are valid in the intervals $0<x<x_m$ and $x_m<x< \lambda$, respectively.
In Eqs.\ (\ref{df}) we have introduced the quantity

\begin{equation} \label{wcrit}
w_*(x)=\sqrt{\psi(x)-\psi_m},
\end{equation}
which has a meaning of the maximal velocity of reflected electrons at
a point $x$.
For the sake of clarity, in Fig.\ \ref{wx} we show the electron trajectories
in the phase space $(x,w)$ corresponding to different electron groups.
It is worth to stress that the distributions (\ref{df}) depend
on the local potential $\psi(x)$ and the potential minimum $\psi_m$ as well,
i.e., the distribution function depends {\em non
locally}
on the potential profile.

Summing up all the contributions (\ref{df}), the total distribution
function takes on the form

\begin{equation} \label{totdf}
f(x,w)=\frac{1}{\sqrt{\pi}} e^{-w^2+\psi(x)} \times
\left\{ \matrix{
e^{-\psi_L}, & w \ge \mp w_*(x), \cr
e^{-\psi_R}, & w < \mp w_*(x). } \right.
\end{equation}
Here, and throughout the paper, we shall use the upper sign for the left
side of the potential minimum $0<x<x_m$ and the lower sign for the right side
of the potential minimum $x_m<x<\lambda$.
It is seen, that the obtained distribution function is discontinuous on $w$
at the points where $w=w_*(x)$ (see also Fig.\ \ref{figdf} discussed below).
It is not surprising, since only a discontinuous solution can satisfy
the first-order equation (\ref{vlasns}) and simultaneously
two different arbitrary functions given at the boundaries.

\subsection{Electron density}

The distribution function (\ref{totdf}) allows us to find the electron density
at a slice $x$ as

\begin{eqnarray} \label{nx}
n(x)&=&\frac{1}{\sqrt{\pi}}e^{\psi(x)}
\left[e^{-\psi_L}\int_{\mp w_*(x)}^{\infty} e^{-w^2} dw \right.
\nonumber\\ &&\hspace{2.5cm} \left.
+ e^{-\psi_R}\int_{-\infty}^{\mp w_*(x)} e^{-w^2} dw\right] \nonumber\\
&=&\frac{1}{2} e^{\psi(x)}
\biglb( e^{-\psi_L} \{ 1 \pm {\rm erf}[w_*(x)] \}
\nonumber\\ &&\hspace{2.5cm}
+ e^{-\psi_R} \{ 1 \mp {\rm erf}[w_*(x)] \} \bigrb),
\end{eqnarray}
where erf$(x)=(2/\sqrt{\pi})\int_0^x e^{-u^2}du$ stands for the error
function. By using the values for the potential (\ref{bcpoisn}) 
at the contacts and denoting

\begin{equation} \label{bet12}
\beta_1 = 1+e^{-V}, \quad \beta_2 = 1-e^{-V},
\end{equation}
the electron density can be written as a function of $\psi$,

\begin{equation} \label{den}
n(\psi)=\frac{1}{2} e^{\psi}\,
[\beta_1 \pm \beta_2 \,{\rm erf}(\sqrt{\psi-\psi_m})],
\end{equation}
where, as before, the upper sign applies in the interval $0<x<x_m$
and the lower sign applies in the interval $x_m<x< \lambda$.
Note that in equilibrium, $V=0$, $\beta_1=2$, $\beta_2=0$, the Boltzmann
distribution $n(x) = e^{\psi(x)}$ is recovered throughout the sample.
Furthermore, Eq.\ (\ref{den}) is valid for a single-injection (vacuum) diode,
assuming $\beta_1=\beta_2=1$. \cite{north40,ziel54}

In the following we shall use the shifted potential measured from the minimum

\begin{equation} \label{etadef}
\eta(x)=\psi(x)-\psi_m,
\end{equation}
and Eq.\ (\ref{den}) in terms of the new variable $\eta$ becomes

\begin{equation} \label{den2}
n(\eta)=n_m e^{\eta}\,
[1 \pm \beta \,{\rm erf}\sqrt{\eta}],
\end{equation}
where $n_m=\frac{1}{2}\beta_1 e^{-V_m}$ is the electron density at
the potential minimum, and

\begin{equation}
\beta\equiv\frac{\beta_2}{\beta_1} = \tanh\left(\frac{V}{2}\right).
\end{equation}

\subsection{Steady-state electrostatic potential}
\label{sspot}

Having found the analytical expression for $n(\psi)$,
we have to use it to solve the Poisson equation (\ref{poisn}).
Multiplying both sides of Eq.\ (\ref{poisn}) by $d\psi/dx$ and
integrating, one gets

\begin{equation} \label{dpsidx}
\left(\frac{d\psi}{dx}\right)^2 =
2 \int_{\psi_m}^{\psi} n(\tilde{\psi}) d\tilde{\psi},
\end{equation}
where we have used the property of the potential minimum
$(d\psi/dx)|_{x=x_m}=0$. Changing to the shifted-potential variable $\eta$
and carrying out the integration, one gets

\begin{equation} \label{poisn2}
\ell_m^2 \left(\frac{d\eta}{dx}\right)^2 = h_{\mp}^V(\eta)
\end{equation}
where $1/\ell_m^2=2n_m=\beta_1 e^{-V_m}$ and the function

\begin{eqnarray} \label{h}
h_V^{\mp}(\eta) = e^{\eta}-1
\pm \beta \left(e^{\eta}{\rm erf}\sqrt{\eta}
- {2\over \sqrt{\pi}}\sqrt{\eta}\right),
\end{eqnarray}
depends on the applied voltage $V$ through $\beta$.
Taking into account $d\eta/dx<0$ for $0<x<x_m$ and $d\eta/dx>0$
for $x_m < x < \lambda$, the electric field is given by

\begin{equation} \label{e}
E = - \frac{d\eta}{dx}=
\left\{ \matrix{
\phantom{+}\sqrt{h_V^-(\eta)}/\ell_m, & 0<x<x_m \cr
-\sqrt{h_V^+(\eta)}/\ell_m, & x_m<x< \lambda } \right.
\end{equation}
which is measured in units of $k_B T/qL_D^0$.
Integrating Eq.\ (\ref{poisn2}), one obtains the distribution of the potential
in an implicit form,

\begin{equation}\label{x} 
x = \left\{ \matrix{ 
\ell_m\int_{\eta}^{\eta_L} 
{\displaystyle \frac{d\eta}{\sqrt{h_V^-(\eta)}} }, &
0<x<x_m \label{xl} \cr
\lambda - \ell_m\int_{\eta}^{\eta_R}
{\displaystyle \frac{d\eta}{\sqrt{h_V^+(\eta)}} }, & 
x_m < x < \lambda \label{xr} } \right.
\end{equation}
where the boundary conditions for $\eta(x)$ are

\begin{equation}
\eta(0) \equiv \eta_L = V_m, \quad \eta(\lambda) \equiv \eta_R = V_m+V.
\end{equation}
For the given $V$,$\lambda$, the only unknown parameter in Eqs.\ (\ref{x})
is the potential minimum $V_m$.
The latter is found by matching Eq.\ (\ref{x}) at $x=x_m$, where $\eta(x_m)=0$,
and one gets

\begin{equation} \label{Vm}
\lambda_m(V) =\int_0^{V_m(V)} \frac{d\eta}{\sqrt{h_V^-(\eta)}}
+ \int_0^{V_m(V)+V} \frac{d\eta}{\sqrt{h_V^+(\eta)}},
\end{equation}
where

\begin{equation} \label{lamm}
\lambda_m=\lambda\sqrt{2n_m}
\end{equation}
is the screening parameter renormalized to the electron density at the
potential minimum rather than to the contact electron density as before.

\subsection{Steady-state current}
This brief description of the steady state is then completed by the expression
for the stationary current.
Substituting the distribution function into Eq.\ (\ref{igen}) and changing
the variables with the help of Eq.\ (\ref{encons}) as $wdw=w_c d w_c$,
one obtains

\begin{eqnarray} \label{j1}
\bar{I} &=& 2\sqrt{\pi}\,I_c \left[ \int_{w^*(x)}^{\infty} f_{L,t}(x,w) w dw
\right. \nonumber\\
&&\hspace{2.5cm} + \left. \int_{-\infty}^{-w^*(x)} f_{R,t}(x,w) w dw \right]
\nonumber\\
&=&2\sqrt{\pi}\,I_c \left[ \int_{w_L}^{\infty} f_{L,t}(0,w_c)\, w_c d w_c
\right. \nonumber\\
&&\hspace{2.5cm} - \left. \int_{w_R}^{\infty} f_{R,t}(\lambda,w_c) \,
w_c d w_c \right],
\end{eqnarray}
where
\begin{equation} \label{ic}
I_c=\frac{1}{\sqrt{\pi}}q N_0 v_0 A=q N_0 \bar{v} A
\end{equation}
is the emission current from each contact (limiting value for the total
current at $V\to\infty$, $V_m\to 0$), and
$\bar{v}=v_0/\sqrt{\pi}=\sqrt{2k_B T/(\pi m)}$ is the average velocity
of the injected electrons with the half-Maxwellian distribution.
Only the part of the distribution function corresponding to the transmitted
electrons has been taken into account, since the reflected carriers gives no
contribution to the current.
[This is in contrast to the case of the calculation of the electron density
(\ref{nx}) for which both transmitted and reflected carriers contribute.]
It is seen from Eq.\ (\ref{j1}) that the current is the same for any section
$x$ of the sample, given by its value at the injected contacts.
Substituting the functions (\ref{bcdf}) into Eq.~(\ref{j1}) and
carrying out the integration, we obtain the current as a sum of two opposing
currents: $I_{LR}$ and $I_{RL}$ caused by the injection from the left and right
contacts, respectively,

\begin{equation} \label{j}
\bar{I} = I_c e^{-V_m} - I_c e^{-V_m - V} \equiv I_{LR} - I_{RL}.
\end{equation}
The formula for the current may be written through the electron density at
the potential minimum, that is,

\begin{equation} \label{jnm}
\bar{I} = 2 n_m I_c \beta = q N_m \bar{v} A \, \tanh \left(\frac{qU}{2k_BT}\right),
\end{equation}
where $N_m$=$2N_0 n_m$.
This formula justifies the usage of the term ``virtual cathode'' referred
to the location of the potential minimum, since it is seen that the current
is determined by the injection of the electron density $N_m$ from the
virtual cathode.
The additional tanh( ) factor takes into account the injection in the
opposite direction, and it tends to 1 at $qU\gg k_B T$.
(For the vacuum diode case, this factor is set to 1 because of only one
injecting contact.)

Summarizing this section, we note that the above relations solve completely
the steady-state problem for the ballistic two-terminal conductor:
Eqs.~(\ref{x}) determine the distribution of the potential
across the diode in an implicit form, and Eqs.~(\ref{Vm}) and (\ref{j})
determine the current-voltage characteristics.
Note that in Eq.\ (\ref{j}) the current depends on voltage through both
the explicit term $e^{-V}$ and the potential minimum $V_m$, which is a
function of voltage.
Equations (\ref{den}), (\ref{h})--(\ref{Vm}), (\ref{j}) may be viewed as
an extension of the Fry-Langmuir theory for a single-injection vacuum
diode \cite{fry,langmuir,ziel-ss76} to the double-injection case.
The Fry-Langmuir formulas are obtained by setting $\beta_1=\beta_2=1$,
$I_{RL}$=0.

\subsection{Results}
\label{res-st}

\begin{figure}
\epsfxsize=8.0cm\epsfbox{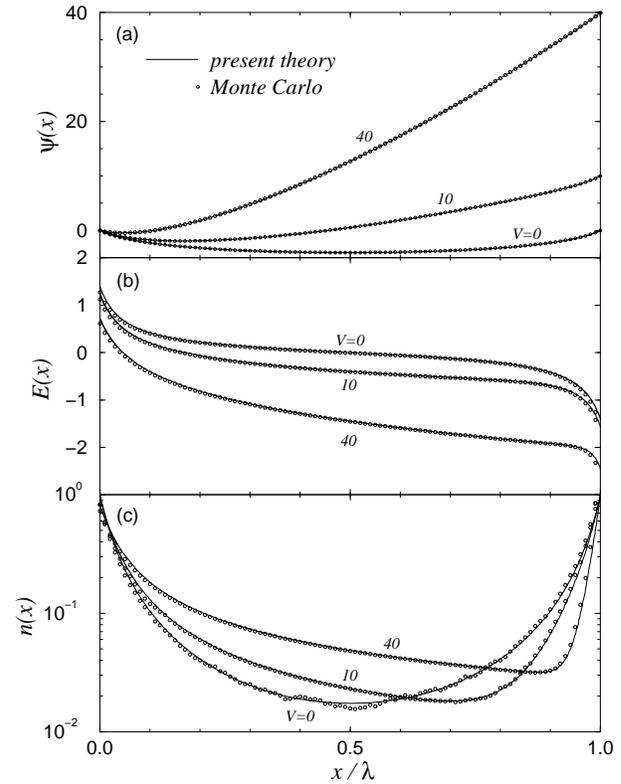}
\protect\vspace{0.5cm}
\narrowtext
\caption{
Spatial profiles for the normalized quantities: (a) potential $\psi$,
(b) electric field $E$, and (c) electron density $n$ (all solid lines)
for $\lambda$=30.9 and several applied biases $V$.
The corresponding units are $k_BT/q$, $k_BT/qL_D^0$, and $2N_0$.
The results are shown to be in excellent agreement with the Monte Carlo
simulations (Ref.\ 48) (symbols).}
\label{figss} \end{figure}

Figure \ref{figss} shows the typical spatial distributions of the potential
$\psi$, electric field $E$, and electron density $n$ along the diode
obtained from Eqs.\ (\ref{den2}), (\ref{e}), (\ref{x}), and (\ref{Vm}).
With the aim to compare our theory with the results of the Monte Carlo
simulations, \cite{gonzalez98b} we present the spatial profiles for the
value of $\lambda$=30.9 and various applied biases $V$.
As it is seen from the figure, the agreement is excellent for all the
quantities.

The space-charge-limited conduction is characterized by a strong transport
inhomogeneity in the ballistic region and by the presence of the potential
minimum [Fig.\ \ref{figss}(a)] due to the injected space charge.
The minimum acts as a barrier for the electrons moving in both directions.
Its magnitude progressively decreases as the applied bias is increased,
simultaneously shifting towards the left contact on which the potential is
fixed.

The obtained solutions are determined by two dimensionless parameters:
$\lambda$ and $V$.
The spatial distributions, however, may be presented in a more universal
form by using for scaling the potential minimum parameters.
We define the new coordinate $\chi=(x-x_m)/\ell_m$,
where the characteristic length $\ell_m=(2n_m)^{-1/2}$,
dependent on the electron density at the potential minimum,
has been introduced in Sec.\ \ref{sspot}.
This is equivalent to scale the original coordinate $X$ in units of
the screening length referred to the electron density at the potential
minimum rather than to the contact electron density.
In such a unit, the parameter $\lambda$ is scaled away from the equation
for the potential, remaining only in the upper and lower bounds
of the function variation. Explicitly, Eq.\ (\ref{x}) becomes

\begin{equation} \label{chi}
\chi =\left\{ \matrix{
-\int_0^{\eta} [d\eta/\sqrt{h_V^-(\eta)}], & -x_m/\ell_m < \chi<0, \cr
\phantom{+}\int_0^{\eta} [d\eta/\sqrt{h_V^+(\eta)]},
& 0<\chi< (\lambda-x_m)/\ell_m. } \right.
\end{equation}
\begin{figure}
\epsfxsize=8.0cm\epsfbox{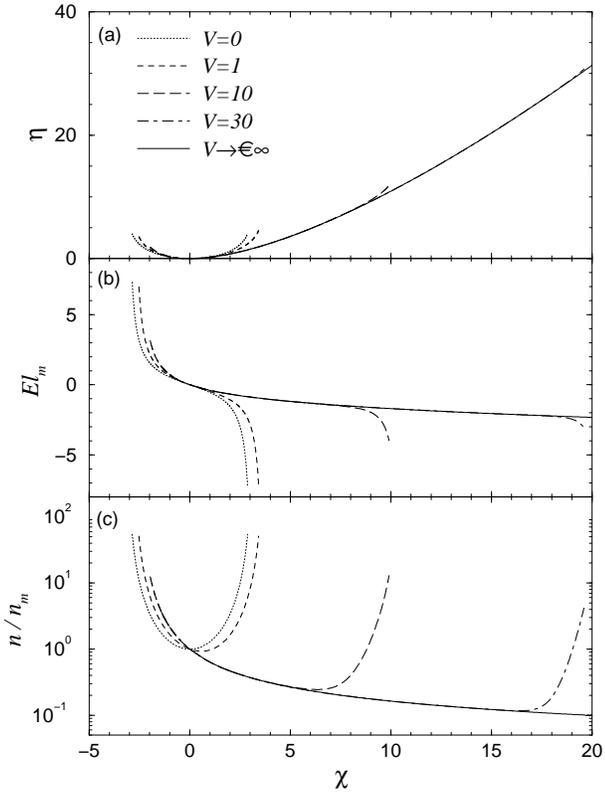}
\protect\vspace{0.5cm}
\narrowtext
\caption{Profiles for the steady-state quantities in units
of the potential-minimum parameters:
(a) potential, (b) electric field, and (c) electron density.
The limiting universal profiles at $V\to\infty$ are shown by solid lines.}
\label{fignor} \end{figure}
\noindent
Therefore, all the solutions may be presented as a one-parameter family
of curves dependent on the applied bias $V$ only.
Moreover, at high-voltage limit $V\gtrsim 5$, $\beta\to 1$, the functions
$h_V^{\pm}(\eta)$ become independent of bias, and the spatial distributions
tend to the limiting {\em universal profiles} for each quantity which are
free from any parameter.
This is valid for all the spatial characteristics as it is seen from
Fig.\ \ref{fignor}, where the potential $\eta=\psi-\psi_m$, the electric
field $E\ell_m$, and the electron density $n/n_m$ are plotted.
Moreover, at this limit the part of each profile at $\chi < 0$ tends
to vanish (the potential minimum approaches the left contact), 
which leads to the validity of the virtual-cathode approximation 
with the boundary condition $E(0)$=0.
The universal profiles obey the asymptotic behavior at $\chi\to\infty$:
$\eta(\chi)=\frac{3}{4} a \chi^{4/3}$,
$E\ell_m = -a \chi^{1/3}$, $n/n_m = \frac{2}{3} a \chi^{-2/3}$,
where $a=(3/\pi)^{1/3}\approx 0.9847$.
Going back from $\chi$ to the $x$ coordinate, and using the Child law
(\ref{ivlchl}), which will be discussed below, one can obtain
the asymptotic formula for the potential profile
$\eta(x)=V(x/\lambda)^{4/3}$, $x\to\infty$.
The latter is valid not only for the present nondegenerate-electron-gas
model, but for an arbitrary distribution function of the injected electrons,
provided $V\to\infty$, $\lambda\to\infty$ (virtual-cathode approximation).
\cite{unpub}
The related formulas for $E(x)$, $n(x)$ may also be obtained by taking
the derivatives.

\begin{figure}
\epsfxsize=8.0cm\epsfbox{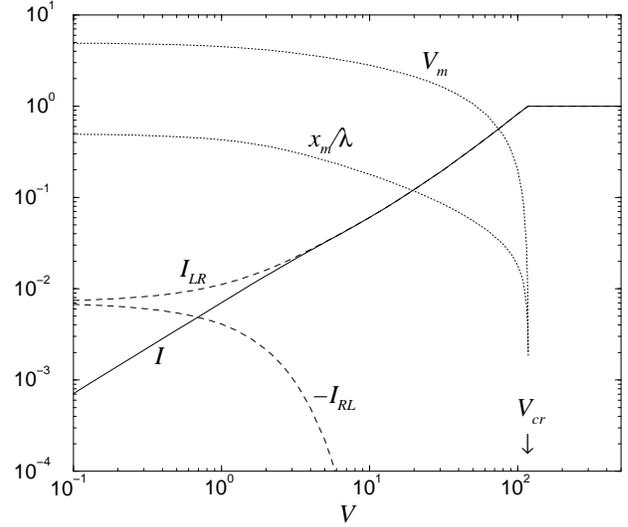}
\protect\vspace{0.5cm}
\caption{Current and its components coming from two opposite electron flows
$\bar{I}=I_{LR}-I_{RL}$ (in units of $I_c$) vs bias $V$ for $\lambda$=50.
The height of the potential barrier $V_m$,
and its location $x_m/\lambda$ are also shown.}
\label{figxm50} \end{figure}

The choice of the potential minimum parameters as
reference coordinates is of traditional use in vacuum-diode literature.
\cite{north40,ziel54,langmuir}
Since only one contact (cathode) is considered as injected for these
diodes, $\beta$=1 for any bias, and the universal potential profile
independent of the diode parameters is obtained for any bias, as it was
tabulated in the original work by Langmuir. \cite{langmuir}
For the case of the two-terminal semiconductor diode, that universality is
broken at low and moderate biases due to the contribution to the current
from the second injecting contact, but it is recovered however at high
biases $V\to\infty$ when the influence of the second contact becomes
negligible, as it is demonstrated in Fig.\ \ref{fignor}.
We remark additionally that the virtual-cathode approximation is only valid
when besides $V\to\infty$ another condition is fulfilled simultaneously,
$V<V_{cr}$.
Otherwise, the transport is no longer limited by the space charge,
the current saturates at $\bar{I}=I_c$, and the value of the electric field
at the left injecting contact is no longer zero, $E(0)<0$.
This change in the transport regime is clearly seen in Fig.\ \ref{figxm50},
where the current and its components coming from two opposite electron flows
$\bar{I}=I_{LR}-I_{RL}$ versus bias $V$ are plotted for a particular value of
$\lambda$.
It is seen that the current is an increasing function of the bias up to
the critical value $V_{cr}$, after which it is saturated at $\bar{I}=I_c$.
At that point the potential minimum vanishes.
For $V\lesssim 5$ the contribution to the current from the right-contact
electrons is also essential.

\begin{figure}
\epsfxsize=8.0cm\epsfbox{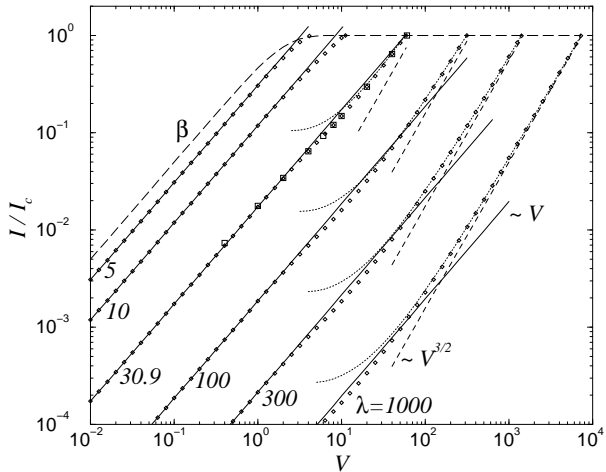}
\protect\vspace{0.5cm}
\caption{
Current-voltage characteristics for different levels of screening $\lambda$
obtained as solutions of the present theory (diamonds).
For comparison, the approximate solutions are shown:
linear dependences given by Eq.\ (\ref{ivlin}) (solid);
Langmuir formula given by Eq.\ (\ref{ivlchl}) (dots);
the Child 3/2-power law given by Eq.\ (\ref{ivlch}) (dashes);
parameter $\beta$, shown by long dashes, when approaching to 1 indicates
the bias ($V\approx 5$) over which the effect of the second contact
on $I$-$V$ curves becomes negligible. Monte Carlo simulation results 
(Ref.\ 48) for $\lambda$=30.9 are shown by squares.}
\label{figiv} \end{figure}

The $I$-$V$ curves for different levels of screening are shown in
Fig.\ \ref{figiv}.
We have checked that the obtained solutions are in excellent agreement with
the Monte Carlo simulations \cite{gonzalez98b}
(the case of $\lambda$=30.9 is compared in the figure).

The analysis shows the following behavior.
At low biases, the $I$-$V$ curves are linear for all $\lambda$ 
despite the fact that the transport is space-charge-limited.
The curves for this case are described by

\begin{equation} \label{ivlin}
I_{\rm lin} \approx I_c \, V e^{-V_m^0}, \qquad  V\lesssim 1,
\end{equation}
where $V_m^0$ is the equilibrium value of the potential minimum whose value
depends on $\lambda$.
In the range $1\lesssim V \lesssim 10$ the $I$-$V$ curves deviate to
sublinear dependence.
At high biases, starting approximately at $V\approx 5$ where $\beta\to 1$,
the effect of injection from the second contact becomes negligible,
$I_{RL}\ll I_{LR}$ (see also Fig.\ \ref{figxm50}).
Furthermore, for $V\gg V_m \gg 1$ the analytical solution may be found.
In this regime, the function $h^+(\eta)$ may be approximated by
leading-order terms of a series expansion in a similar way as in the case
of a vacuum diode \cite{ziel-ss76}

\begin{equation} \label{asseta}
h^+(\eta) \approx 2\sqrt{\eta/\pi} - 1, \qquad \eta\to\infty.
\end{equation}
In this regime from Eq.\ (\ref{xr}) one can write

\begin{equation} \label{etlch}
\lambda - x_m \approx \ell_m \int_{\pi/4}^{V+V_m}
\frac{d\eta}{(2\sqrt{\eta/\pi}-1)^{1/2}},
\end{equation}
from which by using $\ell_m\approx 1/\sqrt{J}$ follows the Langmuir formula
\cite{ziel54,langmuir,ziel-ss76}

\begin{equation} \label{ivlch}
I_{\rm Lang} =
\frac{8}{9} \sqrt{\pi} \,I_c\, \frac{(V+V_m)^{3/2}}{(\lambda-x_m)^2}
\left[ 1 + \frac{3}{\sqrt{\frac{4}{\pi}(V+V_m)}} \right].
\end{equation}
In Fig.\ \ref{figiv} we present the curves calculated from this formula,
and they are seen to describe accurately the $I$-$V$ characteristics for
the highest biases.
For higher $\lambda$, the range of biases where this formula may apply is
wider. In the asymptotic limit $V\to\infty$, $\lambda\to\infty$, one may
neglect $x_m$ and $V_m$ as compared to $\lambda$ and $V$, respectively, and
one obtains the Child 3/2-power law, which is free from the potential minimum
parameters

\begin{equation} \label{ivlchl}
I_{\rm Child} = \frac{8}{9} \sqrt{\pi} \,I_c\, \frac{V^{3/2}}{\lambda^2}.
\end{equation}
It is seen from the figure, that this asymptotic formula accurately describes
the $I$-$V$ curves only at very high values of the parameters:
$\lambda\gtrsim 10^3$, $V\gtrsim 10^3$.
However, as we have discussed earlier, there is a relevant
difference between the semiconductor and vacuum ballistic diodes.
In vacuum diodes the applied voltage may be quite large without
breaking down the ballistic transport regime.
In contrast, in solids, electrons even for a pure material
interact with a lattice.
Under a low-bias regime this interaction is weak, but it becomes quite
strong at high biases due to the significant increase of the electron energy.
For instance, the threshold for the optical phonon generation in GaAs is
about 0.036 eV, which corresponds to $V\approx 40$ at $T\sim 10$K.
Thus, one cannot bias the sample to the voltage more that that value,
since a strong interaction with the lattice will break down the ballistic 
regime. The allowed range of biases is typically restricted by
$U\lesssim 50 k_B T/q$. Then, for real structures the ballistic length
$\lambda$ is well below 100.
Therefore, the Child 3/2-power law is hard to achieve in semiconductor
ballistic $n$-$i$-$n$ diodes,
and one should use the full set of formulas described in the present
paper from which follows the linear or sublinear $I$-$V$ dependences in a wide
range of biases even under a strong limitation of transport by a space charge.

\begin{figure}
\epsfxsize=8.0cm\epsfbox{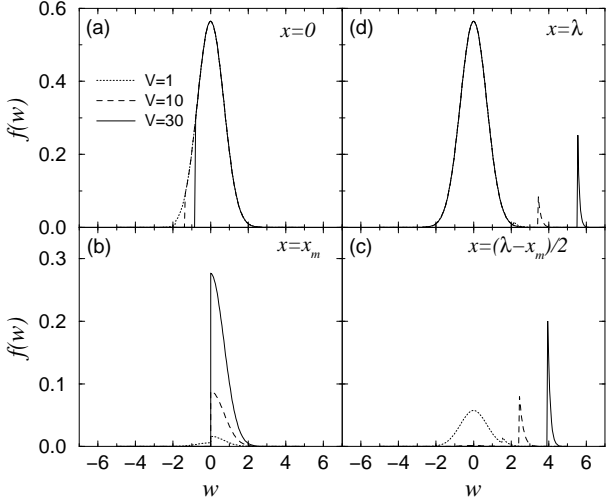}
\protect\vspace{0.5cm}
\caption{
Electron distribution function over dimensionless velocities $f(w)$
at different sections $x$ of the diode for several biases $V$
and $\lambda$=30.}
\label{figdf} \end{figure}

Finally, Fig.\ \ref{figdf} illustrates the stationary electron distribution
function over velocities $f(w)$ at different sections of the diode
for several biases $V$.
The distribution functions are discontinuous at $w=w^*(x)\,{\rm sgn}(x-x_m)$,
as discussed in Sec.\ \ref{stat-df}.
It is interesting to note that at high biases the arriving electrons
at the right (receiving) contact exhibit a sharp peak separated
in energy from the intrinsic contact electrons [Fig.\ \ref{figdf}(d)]
and thus may be distinguished in an experiment. \cite{hayes85,heiblum85}
While the injected carriers are uncorrelated, electrons arriving at 
the receiving contact that belong to that peak exhibit correlations
in energy. This interesting result will be discussed in Sec.\ \ref{seccor}.

\section{Fluctuation problem}

We will find here the fluctuations of the distribution function, electron 
density, electrostatic potential and current in the ballistic region 
of the diode, which are caused by the fluctuations in the contacts.
To solve the fluctuation problem (\ref{vlasf})--(\ref{bcdpois})
self-consistently, we undertake the same approach as used above for
the steady-state problem.
First, we calculate the fluctuation of the distribution function $\delta f$
in a given electrostatic potential $\psi(x)+\delta\psi(x)$ by solving
the perturbed kinetic equation

\begin{equation} \label{kinf}
w \,\frac{\partial\delta f}{\partial x} +
\frac{1}{2} \frac{d\psi}{d x}\, \frac{\partial\delta f}{\partial w}
+\frac{1}{2} \frac{\partial f}{\partial w} \frac{d\delta\psi}{d x}=0.
\end{equation}
The solutions for $\delta f(x,w)$ for different groups of electrons
are derived in the Appendix, where we also present
the corresponding electron density fluctuations $\delta n(x)$ obtained by
integration over velocities.
The fluctuations $\delta n$, which are the functions of $\delta\psi$
and the contact fluctuations $\delta f_L$ and $\delta f_R$, should then be
substituted into the perturbed Poisson equation

\begin{equation} \label{poisf}
\frac{d^2 \delta\psi}{dx^2} = \delta n(x)
\end{equation}
with the boundary conditions

\begin{equation} \label{bcpoisf}
\delta\psi(0) = \delta\psi(\lambda) = 0
\end{equation}
to find the self-consistent fluctuations of the potential $\delta\psi$.

The fluctuations of the distribution functions of injected electrons
$\delta f_L$, $\delta f_R$ are supposed to be given by
the correlator (\ref{corrf}).
It is advantageous, however, to express them through the injected current
fluctuations. For each injection energy $\varepsilon\equiv w_c^2$,
their relationship is given by

\begin{equation} \label{di}
\delta I_k(\varepsilon) = \sqrt{\pi}\,I_c\,\delta f_k(\varepsilon),
\qquad k=L,R,
\end{equation}
where $\delta I_k$ is the low-frequency Fourier component of 
the injection current fluctuation, and 
$I_c$ is the mean emission current defined by Eq.\ (\ref{ic}).
The correlator for $\delta I_k$ is obtained from that for $\delta F_k$
given by Eq.\ (\ref{corrf}), and one gets

\begin{equation} \label{didib}
\langle \delta I_k(\varepsilon)\delta I_{k'}(\varepsilon') \rangle
= 2qI_c\Delta f\,e^{-\varepsilon} \delta_{kk'}
\delta (\varepsilon-\varepsilon'),
\end{equation}
with $\Delta f$ the frequency bandwidth.
The obtained correlator shows that the electrons with different energies
are uncorrelated, which is a consequence of the Poissonian injection
statistics. The fluctuations at the left and right contacts are assumed 
to be uncorrelated as well.

\subsection{Injected electron-density fluctuations}
\label{secinj}

The electron-density fluctuations at a slice $x$ caused by the stochastic
injection from the contacts is obtained by summing up all the contributions
(\ref{dninj}) derived in the Appendix.
In terms of the injected current fluctuations (\ref{di}),
we obtain the following expression:

\begin{eqnarray} \label{dninjtot}
\delta &&n^{inj}(x) = \frac{1}{2\sqrt{\pi}I_c}
\sum_{k=L,R} \int_{\psi_k-\psi_m}^{\infty}
\frac{\delta I_k(\varepsilon)\,
d\epsilon}{\sqrt{\varepsilon+\psi(x)-\psi_k}}
\nonumber\\ &&+ \frac{1}{\sqrt{\pi}I_c} \left\{ \matrix{ {\displaystyle
\int_{\psi_L-\psi(x)}^{\psi_L-\psi_m} \frac{\delta I_L(\varepsilon)\,
d\epsilon}{\sqrt{\varepsilon+\psi(x)-\psi_L}},
} & 0<x<x_m \cr {\displaystyle \int_{\psi_R-\psi(x)}^{\psi_R-\psi_m}
\frac{\delta I_R(\varepsilon)\,
d\epsilon}{\sqrt{\varepsilon+\psi(x)-\psi_R}},
} & x_m<x< \lambda. } \right.
\end{eqnarray}

\subsection{Induced electron-density fluctuations}
\label{secind}

The electron-density fluctuations induced at a slice $x$ by the fluctuations
of the potential is obtained by summing up all the contributions
(\ref{dnind}), and one gets

\begin{eqnarray}\label{dnindtot}
\delta &&n^{ind}(x) = n(x)\,\delta\psi(x) \pm \frac{J}{2\sqrt{\pi}w_*(x)}
[\delta\psi(x)-\delta\psi_m],
\end{eqnarray}
where $J\equiv\bar{I}/I_c$, and
the upper sign applies in the interval $0<x<x_m$
and the lower sign applies in the interval $x_m<x< \lambda$.
This term along with the term (\ref{dninjtot}) should then be used
in the perturbed Poisson equation.

\subsection{Current fluctuations}

The expression for the fluctuation of the current in any section of the sample
is given by

\begin{equation} \label{djgen}
\delta I = 2\sqrt{\pi}\,I_c \int_{-\infty}^{\infty} \delta f(x,w) \, w \,d w.
\end{equation}
Now we have to substitute into Eq.\ (\ref{djgen}) the fluctuation of
the distribution function, which is convenient to consider here as a sum
of the homogeneous and nonhomogeneous parts of the solution of the kinetic
equation [see Eq.\ (\ref{dftot}) in the Appendix].
The contribution of the nonhomogeneous term is zero, which can be easily
checked by direct integration of Eq.\ (\ref{ddfnhom1}).
The homogeneous term consists of the transmitted and reflected parts
given by Eqs.\ (\ref{ddfhom}).
Again, the reflected electrons give zero contribution to the current
fluctuations, since the functions (\ref{ddfhomb}) are even on $w$,
so that the integrand (\ref{djgen}) is an odd function
and its integration from $-\infty$ to $\infty$ yields zero.
The only nonzero contribution comes from the terms (\ref{ddfhoma})
for transmitted perturbing electrons.
Substituting them into Eq.\ (\ref{djgen}) and changing the variable
of integration from $w$ to $w_c$, we obtain

\begin{eqnarray} \label{dj1}
\delta I &=& 2\sqrt{\pi} \,I_c \left[
\int_{w_L}^{\infty} \delta f_L(w_c)\,w_c\, d w_c \right. \nonumber\\
&&\hspace{1cm}
+ \left.\int_{-\infty}^{-w_R} \delta f_R(w_c)\,w_c\, d w_c \right]
\nonumber\\ &&\hspace{1cm}
+ I_c\, e^{\psi_m} (e^{-\psi_L} - e^{-\psi_R}) \delta \psi_m
\end{eqnarray}
from which it is seen that the current fluctuation is independent of the
position $x$.
By using the definition for the injected current fluctuation
(\ref{di}) and the formula (\ref{j}) for the average current,
the final expression for the current fluctuation takes on the form

\begin{eqnarray} \label{dj}
\delta I =\int_{V_m}^{\infty} \delta I_L(\varepsilon)\,d\varepsilon
- \int_{V_m+V}^{\infty} \delta I_R(\varepsilon) \,d\varepsilon
- \bar{I} \delta V_m,
\end{eqnarray}
where $\delta V_m\equiv -\delta\psi(x_m)$ is the potential minimum fluctuation.
$\delta I$ depends on the magnitude of the fluctuating potential barrier 
irrespective of its random location.
This is a consequence of the current conservation along the diode.

Equation (\ref{dj}) is a central one, which determines the fluctuation of
the transmitted current through the fluctuations injected from the contacts.
The first two terms in the rhs represent the current fluctuations transmitted
directly to the opposite contact from the left and right contacts,
respectively. Since the injected electrons of different energies
are uncorrelated, they give the full shot noise.
It is the last term $-\bar{I}\delta V_m$, caused by the self-consistent 
potential fluctuation (long-range Coulomb correlations),
that compensates the current fluctuation and may result in the noise reduction.
We note, first, that it is proportional to the current and thus exists
only under nonequilibrium conditions.
Second, it depends on the potential barrier fluctuation $\delta V_m$.
When the barrier does not appear under certain conditions,
all the injected fluctuations are transmitted to the opposite contact
and the noise of the transmitted current is expected to be the same
as thet for the injected carriers, i.e., the full Poissonian shot noise.
The compensating behavior may occur only when the potential barrier
is present.
Notice that the contributions of the left- and right-injected
fluctuations are of the opposite sign, i.e., $\delta I_L>0$ increases 
the fluctuation of the transmitted current, while $\delta I_R>0$ decreases it.

Among all the injecting perturbing electrons, only those able to pass over
the potential barrier contribute to the transmitted current fluctuation.
This fact is reflected in the lower integration limits that contain
the height of the potential barrier.
In contrast, all the injected electrons contribute to the potential
barrier fluctuations, and thereby participate in the compensation effect,
as it will be shown in the next section.

\subsection{Self-consistent potential fluctuations}

We find the potential barrier fluctuation $\delta V_m$, which is 
of prime interest, from the linearized Poisson equation (\ref{poisf}) 
for the potential fluctuations $\delta\psi$.
By substituting the electron-density fluctuations $\delta n$ consisting of
the injected and induced contributions found in Secs.\ \ref{secinj} and
\ref{secind}, we obtain for the self-consistent potential fluctuations

\begin{eqnarray} \label{poisp}
\frac{d^2 \delta\psi}{dx^2} &=& \delta n^{ind}(x) + \delta n^{inj}(x)
\nonumber\\
&=& n(x)\,\delta\psi(x) \pm \frac{J}{2\sqrt{\pi}w_*(x)}\,
[\delta\psi(x) - \delta\psi_m] \nonumber\\
&& + \delta n^{inj}(x).
\end{eqnarray}
This is a second-order nonhomogeneous differential
equation with spatially dependent coefficients,
where the term $\delta n_x^{inj}$, dependent on the fluctuations
at the contacts $\delta I_k$ [see Eq.\ (\ref{dninjtot})],
plays the role of a stochastic noise source.
To find its solution in a general form is a complicated problem.
In addition, we remark that the term with $1/w_*(x)$ is singular
at the potential barrier minimum $x=x_m$ which produces an additional
difficulty.
Nevertheless, we will show that it can be solved exactly without any
approximation.
First of all, it is advantageous to introduce a new stochastic quantity

\begin{equation} \label{etax}
\delta\eta_x=\delta\psi(x)-\delta\psi_m,
\end{equation}
which is the the potential fluctuation at a slice $x$ measured from
the {\em fluctuating} potential minimum.
Thus, due to our choice, at the potential minimum $\delta\eta_{x_m}$=0,
where $x_m=x_m^0+\delta x_m$ is a stochastic location of the
potential minimum fluctuating around its steady-state position $x_m^0$.
The latter fluctuation, however, may be neglected, since it is only of
second order in respect to the potential fluctuations,
because of the property of the minimum $\psi'(x_m^0)=0$.
Thus, one gets the stochastic differential equation

\begin{eqnarray} \label{nhom}
\hat{L}\delta\eta_x&\equiv& \left[\frac{d^2}{dx^2}-n(x)
\mp\frac{J}{\sqrt{4\pi\eta(x)}} \right] \delta\eta_x \nonumber\\
&&\hspace{2cm} = -n(x)\delta\eta_L + \delta n^{inj}(x),
\end{eqnarray}
The boundary conditions for this equation follows from Eqs.\ (\ref{bcpoisf})
and (\ref{etax})

\begin{equation} \label{bceta}
\delta\eta_L=\delta\eta_R=-\delta\psi_m.
\end{equation}
Since the potential $\delta\eta_x$ is referenced to the fluctuating minimum,
its values on the contacts are not zero, while in a stationary frame
they are zero due to a fixed-applied-voltage conditions.

To find the solution of Eq.\ (\ref{nhom}), we use a method we have recently
applied for a stochastic drift-diffusion equation which has a similar form.
\cite{apl97}
Essentially, this method is based on the possibility of finding two (arbitrary)
linearly independent solutions of the corresponding homogeneous equation
$\hat{L}\delta\eta_x=0$,
which can further be used to construct the solution for the nonhomogeneous
equation satisfying the appropriate boundary conditions.
One of the solutions is proportional to $(d\psi/dx)$,
which can be seen by differentiating the Poisson equation (\ref{poisn})
and comparing the result with Eq.\ (\ref{nhom}) with zero rhs.
For convenience, we take it as $E(x)=-(d\psi/dx)$,
so the solution coincides with the electric field profile.
In general, \cite{apl97} the second solution can be obtained from the first
one by using the formula $u(x)=E(x)\int_C^x [W(y)/E^2(y)]dy$,
where $W(x)=E(x)u'(x)-E'(x)u(x)$ is the Wronskian,
$C$ is an arbitrary constant, the prime stands for the derivative,
and $E(x)\neq 0$, $\forall x$ is assumed.
However, this formula cannot be applied for our problem, since $E(x)$=0
precisely at the point of the potential minimum and the integral diverges.
Alternatively, we use another formula for the second solution $u$ which
has no divergence in the whole region. Explicitly,

\begin{equation} \label{uc}
u(x) = -\frac{W(x)}{E'(x)} + E(x) \int_C^x \frac{W(y)Q(y)}{[E'(y)]^2}\,dy,
\end{equation}
where the function $Q(x)=-n(x)\mp J/\sqrt{4\pi\eta(x)}$
is a free term in the operator $\hat{L}$,
and the necessary condition $E'(x)\neq 0$ is fulfilled.
Next we notice that the differential operator $\hat{L}$ given by
Eq.\ (\ref{nhom}) does not contain the term with the first derivative,
which leads to the constant Wronskian.
The value of this constant is not actually important, since it will be
canceled as will be seen below, so we take $W(x)=1$. The arbitrary 
constant $C$ in Eq.\ (\ref{uc}) does not influence on the final results.
It is convenient, however, to define it by
the conditions $u(0)$=$u(\lambda)$=0 at the ends of the diode,
which correspond to the homogeneous boundary conditions for
the Green functions of the operator $\hat{L}$ and provide
the most compact intermediate expressions.
To satisfy the zero boundary conditions on both ends of the diode,
one can take the function $u(x)$ as consisting of two branches.
As a result, we obtain the following expression:

\begin{eqnarray} \label{u}
u&&(x)= \frac{1}{n(x)} + E(x) \nonumber\\ && \times \left\{ \matrix{
\displaystyle{
\int_0^x \frac{J\nu(y)+n(y)}{n^2(y)}\,dy -\frac{1}{n_L E_L}, }
& 0 < x < x_m \cr \displaystyle{
\int_x^{\lambda}\frac{J\nu(y)-n(y)}{n^2(y)}\,dy -\frac{1}{n_R E_R}, }
& x_m < x < \lambda, } \right.
\end{eqnarray}
where $\nu(x)\equiv 1/\sqrt{4\pi\eta(x)}$ and
$n(x)$ and $E(x)$ are the steady-state spatial profiles of the electron
density and electric field, which take the values at the left and right
contacts $n_L,E_L$ and $n_R,E_R$, respectively.
The function $u(x)\ge 0$ is continuous in the entire region $0<x<\lambda$, 
including the point of the potential minimum,
where it takes the value $u(x_m)=1/n_m$.
At that point, however, it has an infinite derivative, which is a consequence
of the zero of the field.

\begin{figure}
\epsfxsize=7.0cm\epsfbox{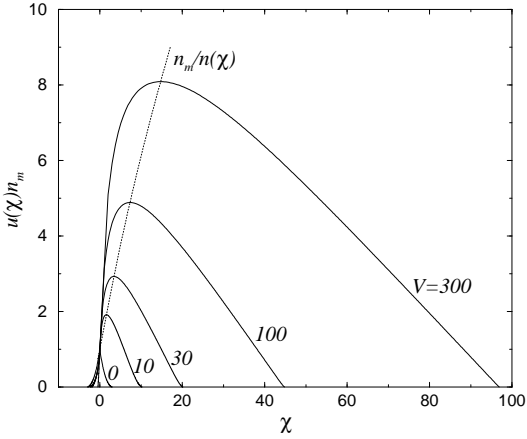}
\protect\vspace{0.5cm}
\caption{Function $u(\chi)$ which shows the relative contributions of the
noise sources $\delta n^{inj}(\chi)$ to the potential barrier fluctuations
for $\lambda$=100 and for several biases $V$.
The potential minimum is located at $\chi$=0.}
\label{figu} \end{figure}

The general solution of Eq.\ (\ref{nhom}), satisfying the boundary conditions
(\ref{bceta}) and the conditions $E(x_m)$=0, $\delta\eta_{x_m}$=0, then reads

\begin{mathletters} \label{detax} \begin{eqnarray}
\delta\eta_x &=& E(x) \int_0^x u(y) \delta s_y\,dy
+ u(x) \int_x^{x_m} E(y) \delta s_y\,dy \nonumber\\
&& + \delta\eta_L \frac{E(x)}{E_L}, \quad 0 < x < x_m, \\
\delta\eta_x
&=& -E(x) \int_x^{\lambda} u(y) \delta s_y\,dy
- u(x) \int_{x_m}^x E(y) \delta s_y\,dy \nonumber\\
&& + \delta\eta_L \frac{E(x)}{E_R}, \quad x_m < x < \lambda,
\end{eqnarray}\end{mathletters}
where $\delta s_x= n(x) \delta\eta_L - \delta n^{inj}(x)$ is
the nonhomogeneous part of Eq.\ (\ref{nhom}).
Thus, one can find the potential fluctuation $\delta\eta_x$ at any section
$x$ of the sample. In particular, its value at the boundaries
yields the potential barrier fluctuation $\delta V_m=\delta\eta_L$.
We find the unknown $\delta\eta_L$ from the continuity condition on
the derivative $d\delta\eta/dx$ at $x=x_m$:

\begin{equation}
\delta\eta_L \left[\frac{1}{E_R}-\frac{1}{E_L}\right] =
\int_0^{\lambda} u(x) \delta s_x\,dx.
\end{equation}
Now recalling that $\delta\eta_L$ has entered also in $\delta s$,
we obtain

\begin{equation} \label{dvmd}
\delta V_m = \frac{1}{\Delta}
\int_0^{\lambda} u(x) \delta n^{inj}(x)\,dx,
\end{equation}
with

\begin{equation}
\Delta = \frac{1}{E_L}-\frac{1}{E_R}
+\int_0^{\lambda} u(x) n(x)\,dx.
\end{equation}
The last integral can further be reduced by substituting $n=-dE/dx$
and the expression for $u(x)$ given by Eq.\ (\ref{u}).
Integrating by parts, one gets the simple formula

\begin{equation} \label{Delt}
\Delta = \frac{\lambda}{2} + \frac{1}{E_L} - \frac{1}{E_R}.
\end{equation}
The obtained analytical expression (\ref{dvmd}) with the parameter $\Delta$
given by Eq.\ (\ref{Delt}) yields the fluctuation of the barrier height 
in terms of the spatially distributed ``noise source'' $\delta n^{inj}(x)$
caused by the random injection from the contacts.
The weight function $u(x)$ shows the relative contributions of
the ``noise sources'' to the potential barrier fluctuations.
Its behavior is illustrated in Fig.\ \ref{figu}, where we present $u$
normalized to $1/n_m$ as a function of the coordinate $\chi=(x-x_m)/\ell_m$.
In such a scaling for a fixed voltage, $u(\chi)n_m$ is almost independent
of $\lambda$ with a slight deviation at the ends of the function extension.
An interesting property of those functions for different biases is that
they cross the curve $n_m/n(\chi)$ (the inverse universal density
profile as discussed in Sec.\ \ref{res-st}) at two characteristic points:
the potential minimum $\chi$=0 where $du/d\chi=\infty$, and at the maximum
of $u(\chi)$ (see Fig.\ \ref{figu}). 
The latter point has significance in that the electron-density fluctuations 
there have the largest influence on the potential-barrier fluctuations.
It is worth noting that the maximum contribution to $\delta V_m$ does not
come from the potential minimum location, as it would seem intuitively.

\subsection{Current noise spectral density}

Substituting the obtained formula (\ref{dvmd}) for $\delta V_m$ into
Eq.\ (\ref{dj}), we obtain the current fluctuation as

\begin{eqnarray}
&&\delta I = \int_0^{\infty}
\gamma_L(\varepsilon)\delta I_L(\varepsilon)\,d\varepsilon +
\int_0^{\infty}\gamma_R(\varepsilon)\delta I_R(\varepsilon)\,d\varepsilon,
\label{dJg} \\
&&\gamma_L(\varepsilon) = \left\{ \matrix{
-2 J \int_0^{x_L^*} K(x,\varepsilon)\,dx, & \varepsilon < V_m, \cr
1 - J \int_0^{\lambda}
K(x,\varepsilon)\,dx, & \varepsilon > V_m, } \right. \label{gam1} \\
&&\gamma_R(\varepsilon) = \left\{ \matrix{
-2J\int_{x_R^*}^{\lambda} K(x,\varepsilon-V)\,dx, & \varepsilon<V_m+V, \cr
-1 - J \int_0^{\lambda}
K(x,\varepsilon-V)\,dx, & \varepsilon > V_m+V, } \right. \label{gam2}
\end{eqnarray}
where
$K(x,\varepsilon)=u(x)/[2\sqrt{\pi}\Delta\sqrt{\varepsilon+\psi(x)}]$,
and $x_L^*$ and $x_R^*$ are found from
$\varepsilon$=$-\psi(x_L^*)$=$V-\psi(x_R^*)$.
The functions $\gamma_k(\varepsilon)$ introduced for each contact
have the meaning of {\em current fluctuation transfer functions},
since they represent the ratio of the transmitted current fluctuation to the
injected current fluctuation for a particular injection energy $\varepsilon$.
The terms proportional to the current $J$ originate from
the potential minimum fluctuations, whereas the constant contributions
($\pm 1$) represent the direct transmission of fluctuations
to the opposite contact.

Equation (\ref{dJg}) leads to the spectral density of current fluctuations

\begin{equation} \label{SI}
S_I = 2qI_c \int_0^{\infty} \left[ \gamma_L^2(\varepsilon)
+ \gamma_R^2(\varepsilon) \right] e^{-\varepsilon}\,d\varepsilon.
\end{equation}
This equation with $\gamma_k(\varepsilon)$ given by formulas
(\ref{gam1}) and (\ref{gam2}) is the final result of our derivations.
It allows us to obtain the current-noise spectral density, for the given
level of screening $\lambda$ and applied voltage $V$, from
{\em the steady-state distributions} of the potential $\psi(x)$, electric
field $E(x)$, and electron density $n(x)$ by direct integration.
Thus, the current-noise level is directly related to the transport
inhomogeneity in the system.
Note that the obtained formulas are exact for biases ranging from
thermal to shot-noise limits under a space-charge-limited transport
conditions.

For practical calculations of the transfer functions
$\gamma_k(\varepsilon)$, one may integrate by parts the function $K$ in
formulas (\ref{gam1}) and (\ref{gam2}), which leads to the following
expressions corresponding to each group of carriers:

\begin{mathletters}\begin{eqnarray}
\gamma_{L,r}(\tilde{\varepsilon}) &=&
- \frac{\beta}{2\Delta_m} \int_{-\tilde{\varepsilon}}^{\eta_L}
\frac{G(\eta,\tilde{\varepsilon})}{[h_V^-(\eta)]^{3/2}}d\eta,
\quad \tilde{\varepsilon} < 0, \label{gamlr} \\
\gamma_{L,t}(\tilde{\varepsilon}) &=&
1 - \frac{\beta}{2\Delta_m} \left\{ \int_0^{\eta_L}
\frac{H(\eta,\tilde{\varepsilon})}{[h_V^-(\eta)]^{3/2}}d\eta \right.
\nonumber\\ && \hspace{1cm}
+ \left. \int_0^{\eta_R}
\frac{H(\eta,\tilde{\varepsilon})}{[h_V^+(\eta)]^{3/2}}d\eta \right\},
\quad \tilde{\varepsilon} > 0, \label{gamlt} \\
\gamma_{R,r}(\tilde{\varepsilon}) &=&
- \frac{\beta}{2\Delta_m} \int_{V-\tilde{\varepsilon}}^{\eta_R}
\frac{G(\eta,\tilde{\varepsilon}-V)}{[h_V^+(\eta)]^{3/2}}d\eta,
\quad \tilde{\varepsilon} < 0, \label{gamrr} \\
\gamma_{R,t}(\tilde{\varepsilon}) &=&
\gamma_{L,t}(\tilde{\varepsilon}-V)-2, \quad \tilde{\varepsilon} > 0,
\label{gamrt}
\end{eqnarray} \end{mathletters}
where $\tilde{\varepsilon}=\varepsilon-V_m$ is the injection electron energy
referenced from the potential minimum,

\begin{eqnarray}
\Delta_m \equiv \frac{\Delta}{\ell_m} &=& \frac{\lambda_m}{2} +
\frac{1}{\sqrt{h_V^-(\eta_L)}} + \frac{1}{\sqrt{h_V^+(\eta_R)}}, \\
H(\eta,\tilde{\varepsilon})&\equiv&\frac{2}{\sqrt{\pi}}
[\sqrt{\eta + \tilde{\varepsilon}}-\sqrt{\tilde{\varepsilon}}], \\
G(\eta,\tilde{\varepsilon})&\equiv&\frac{4}{\sqrt{\pi}}
\sqrt{\eta + \tilde{\varepsilon}}.
\end{eqnarray}
Formulas (\ref{gamlr}) and (\ref{gamlt}) with $\beta$=1 correspond
to the formulas for a vacuum diode found by North within different approach
[see Eqs.\ (31) and (38) of Ref.\ \onlinecite{north40}].

\subsection{Nyquist equilibrium noise}
\label{eqnoise}

In equilibrium, $\bar{I}\to 0$, the compensating term $\bar{I}\delta V_m$ in
Eq.\ (\ref{dj}) vanishes, and, comparing with Eq.\ (\ref{dJg}),
the transfer functions are simply
the step functions with a step at the barrier height:
$\gamma_L^{eq}(\varepsilon)=\theta(\varepsilon-V_m)$,
$\gamma_R^{eq}(\varepsilon)=-\theta(\varepsilon-V_m)$.
This means that only electrons able to pass over the barrier contribute
to the equilibrium (thermal) noise.
For this case, one can easily obtain the Nyquist noise formula

\begin{equation}
S_I^{eq} = 4qI_c e^{-V_m^0} = 4k_BT\,g_0,
\end{equation}
where $g_0=d\bar{I}/dU|_{U\to 0}$ is the zero-bias small-signal conductance.
[To find the conductance we have made use of Eq.\ (\ref{j}).]
Both electron streams, from the left and right contacts, equally contribute
to the Nyquist noise. The space-charge effect on the equilibrium noise
is present in the dependence of $g_0$ on the potential minimum $V_m$.

\subsection{Noise-reduction factor}

The obtained formula (\ref{dvmd}) for the current-noise spectral density
$S_I$, which accounts for the long-range Coulomb correlations, may be compared
with the uncorrelated value through the so-called noise reduction factor.
Out of equilibrium, if one neglects the term $\bar{I}\delta V_m$ in 
Eq.\ (\ref{dj}), which is responsible for the long-range Coulomb correlations 
between the carriers, one obtains
$\gamma_L^{uncor}(\varepsilon)=\theta(\varepsilon-V_m)$,
$\gamma_R^{uncor}(\varepsilon)=-\theta(\varepsilon-V_m-V)$, which leads to

\begin{eqnarray}
S_I^{uncor} = 2q\,(I_{LR}+I_{RL}) &=& 2q\bar{I}\coth(V/2) \\
&\approx& 2q\bar{I}, \quad V\gtrsim 5, \nonumber
\end{eqnarray}
which is nothing more than the Poissonian noise of two uncorrelated streams
of carriers opposite each other (at high voltages the contribution from
the right-contact stream becomes negligible).
It is reasonable, therefore, to define the noise-reduction factor by

\begin{equation} \label{Gamma}
\Gamma =\frac{S_I}{S_I^{uncor}} = \frac{S_I}{2q\bar{I}\coth(V/2)}.
\end{equation}
By this definition, both the thermal noise and shot noise limits are included.
\cite{gonzalez97}

\begin{figure}
\epsfxsize=8.0cm\epsfbox{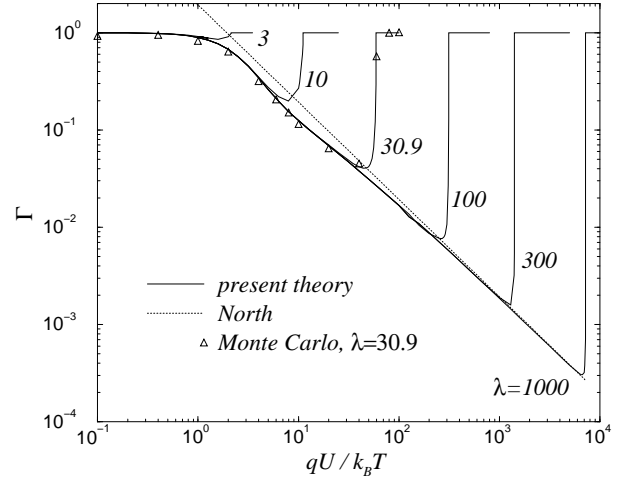}
\protect\vspace{0.5cm}
\caption{Current-noise reduction factor $\Gamma$ vs bias $U$
for different levels of screening $\lambda$=$d/L_D$ (solid).
For comparison, North's asymptotic solution given by Eq.\ (\ref{gnorth})
is shown (dots).
For the case of $\lambda$=30.9, the results are shown to be in excellent
agreement with the Monte Carlo simulations (Ref.\ 5) (triangles).}
\label{figg} \end{figure}

Figure \ref{figg} shows $\Gamma$ versus applied voltage $V$ for various 
screening parameters $\lambda$.
At low values of $\lambda$, the noise-reduction effect is weak,
$\Gamma\approx 1$.
As $\lambda$ increases, the noise becomes substantially reduced
in the range of biases $k_B T\lesssim qU < qU_{cr}$, where $U_{cr}$ is
a critical voltage for which the potential minimum vanishes
(its value is a function of $\lambda$).
At $U\ge U_{cr}$ the full shot-noise level is abruptly recovered.
This sharp increase in the noise intensity when observed in an experiment
would indicate on the disappearance of the potential barrier controlling
the current.

\begin{figure}
\epsfxsize=7.0cm\epsfbox{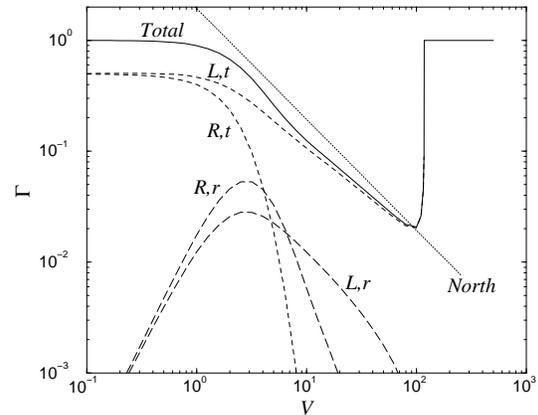}
\protect\vspace{0.5cm}
\caption{
Contributions to the current-noise-reduction factor $\Gamma$ corresponding
to different electron groups for the case of $\lambda$=50.
$L$ and $R$ refer to the left and right contacts, and
$t$ and $r$ distinguish the transmitted and reflected groups of carriers.
North's asymptotic solution is shown by dots.}
\label{Gcon} \end{figure}

We have compared our results for the noise reduction factor with those
obtained by the Monte Carlo simulations. \cite{gonzalez97}
The agreement was found to be perfect within numerical uncertainty of the
Monte Carlo algorithm, as it is seen from Fig.\ \ref{figg}
where we show such a comparison for $\lambda$=30.9.
The agreement for the noise characteristics, as well as for the steady-state
spatial profiles and $I$-$V$ curves, indicates the correspondence between
our kinetic theory and the Monte Carlo model used in
Refs.\ \onlinecite{gonzalez97,b98} and \onlinecite{gonzalez98b}.

An advantage of our analytical approach is that, in addition to the net
noise characteristics, one may distinguish the relative contributions
to the noise from different groups of carriers.
In Fig.\ \ref{Gcon} we present the results for the noise-reduction factor
$\Gamma$ as a sum of four contributions.
It is seen that in equilibrium only the transmitted electrons contribute
to the noise (equally from the left and right contacts).
In the range $1\lesssim V \lesssim 10$, the contribution from the reflected
carriers becomes appreciable with a maximum at $V\approx 3$.
At higher voltages, as the potential barrier progressively decreases,
the role of the reflected carriers becomes less important.
The contribution of the right-contact transmitted electrons is negligible
at $V\gtrsim 5$, as for the stationary $I$-$V$ characteristics.
As a result, in the high-voltage limit, only the left-contact transmitted
electrons contribute to the noise.
This fact can be taken into account in analyzing the asymptotic behavior
of the noise-reduction factor at high-voltage limit.
In this limit the main contribution to the current fluctuation transfer
function comes from $\gamma_{L,t}$.
Under the condition $V_m\ll V <V_{cr}$, which is easy to satisfy at large
$\lambda$,
the first integral in Eq.\ (\ref{gamlt}) is much less than the second one,
so that the contribution to the noise from the region before the virtual
cathode may be neglected. Furthermore, at sufficiently high $\lambda$,
$\Delta_m\approx{1\over 2}\lambda/\ell_m\approx{1\over 2}\lambda\sqrt{J}$.
Thus, one can write

\begin{eqnarray}
\gamma&&_{L,t}(\tilde{\varepsilon}) \approx
1 - \frac{1}{\lambda\sqrt{J}} \int_0^V
\frac{H(\eta,\tilde{\varepsilon})}{[h_V^+(\eta)]^{3/2}}d\eta \nonumber\\
&&\approx 1 - \frac{\pi^{1/4} 2^{3/2} V^{3/4}}{\lambda\sqrt{J}}
\left[ {1\over 3} +
\left( {3\over 4}\sqrt{\pi}-\sqrt{\tilde{\varepsilon}} \right) V^{-1/2}
\right],
\end{eqnarray}
where we have taken into account that the main contribution comes
at the upper integration limit and made use of the asymptotic expansion
of the function $h_V^+$ given by Eq.\ (\ref{asseta}).
It is also assumed here that for any fixed energy the bias is high,
$V\gg\tilde{\varepsilon}$. It is justified since the range of valuable
energies is limited by the Maxwellian exponentially decaying distribution.
Now, substituting the Langmuir expression (\ref{ivlch}) for the current
and neglecting $x_m$ and $V_m$, one obtains

\begin{equation}
\gamma_{L,t}(\tilde{\varepsilon}) \approx \frac{3}{\sqrt{V}}
\left( \sqrt{\tilde{\varepsilon}} - \frac{\sqrt{\pi}}{2} \right),
\end{equation}
This formula, after the integration over the energies, leads to North's
asymptotic formula \cite{north40} for the noise-reduction factor:

\begin{equation} \label{gnorth}
\Gamma \approx \frac{9}{V} \left( 1 - \frac{\pi}{4} \right)
\approx \frac{1.9314}{V},  \qquad V\to\infty.
\end{equation}
This formula is universal in the sense that it is free from any diode
parameter including the screening parameter $\lambda$.
However, it is assumed that $\lambda$ should be sufficiently high to satisfy
the simultaneous conditions $V\to\infty$ and $V < V_{cr}$.
As it is seen from Fig.\ \ref{figg}, the noise-reduction factor $\Gamma$
approaches this asymptotic formula  at high values of the parameters:
$\lambda\gtrsim 10^3$, $V\gtrsim 10^3$.
As we have already noted, in semiconductors it is hard to maintain
the ballistic regime at biases $V\gtrsim 50$ because of
the increasing significance of electron-phonon interactions
which destroy the ballistic regime.
In the range of interest $1\lesssim V \lesssim 50$ the noise level is seen
to be significantly lower than North's asymptotic curve.
This means that the full set of formulas are necessary to describe properly
the noise intensity in the semiconductor ballistic diodes.
Another important conclusion from Fig.\ \ref{figg} is that
for a nondegenerate electron gas there exists the lowest noise-reduction
level dependent only on the bias and the temperature through the factor 
$qU/(k_BT)$, and it is impossible to surmount it by any choice of 
the material parameter and/or geometrical parameters of the diode.
This universal minimal-noise curve approaches North's asymptotic curve
at high voltages.

\subsection{Spectroscopy of shot noise}
\label{seccor}

\begin{figure}
\epsfxsize=8.0cm\epsfbox{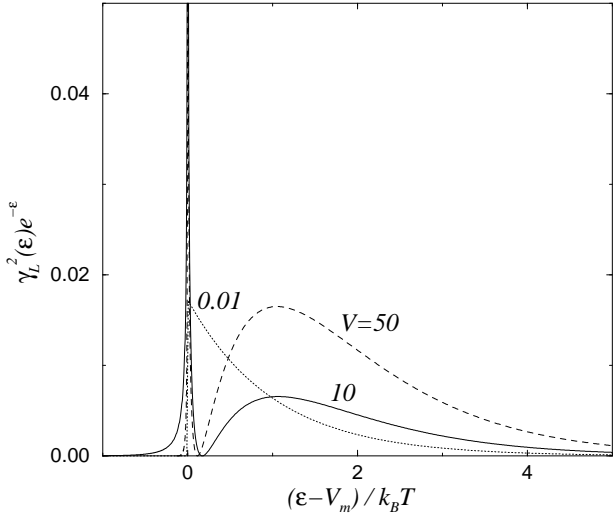}
\protect\vspace{0.5cm}
\caption{
Partial contributions to the current-noise spectral density from different
energies $\varepsilon$ of electrons injected from the left contact for
biases $V$=0.01; 10; 50.
The results for the right-contact electrons are approximately the same for
$V$=0.01 and negligibly  small for $V$=10 and 50.}
\label{g2e} \end{figure}

A great advantage of the derived formula (\ref{SI}) for the current-noise
spectral density is that one may obtain the partial contribution to the noise
from electrons of different injection energies by computing the current
fluctuation transfer functions $\gamma_k(\varepsilon)$.
The electrons for which $\gamma_k(\varepsilon)<0$ reduce the current
fluctuations.
For instance, the right-contact electrons always reduce them, since
$\gamma_R(\varepsilon)<0$, $\forall\varepsilon$.
The reflected carriers originated from the left contact ($\varepsilon<V_m$)
also provide negative values for the transfer function and compensate
the current fluctuations by virtue of the potential-barrier fluctuations.
The same effect is produced by the left-contact transmitted electrons
with the energies slightly above the barrier height $V_m$.
{}From both groups, the most efficient compensation carriers
are those with the energies in the vicinity of $V_m$
where $\gamma_L\to -\infty$. \cite{remark3}
They provide an over-compensation of the injected
from the contacts fluctuations.
In contrast, the injected electrons whose energy greatly exceeds $V_m$
produce negligible perturbations of the potential
barrier, thus leading to the asymptotic behavior
$\gamma_L(\varepsilon)\to 1$, $\gamma_R(\varepsilon)\to -1$
as $\varepsilon\to\infty$.
There also exists the specific energy $\varepsilon^*$,
for which the compensation fluctuation is exactly equal to the injected
fluctuation, giving no noise at all, $\gamma_L(\varepsilon^*)$=0.
This curious fact is illustrated in Fig.\ \ref{g2e} where we present
the contribution to the current-noise spectral density from different
energies of electrons injected from the left contact
by plotting the function $\gamma_L^2(\varepsilon)e^{-\varepsilon}$.
At high biases, just after the peak at $\varepsilon=V_m$,
the point with zero contribution to the noise is observed.
While at equilibrium the maximum contribution comes from the carriers
injected with $\varepsilon=V_m$, at high biases, when the noise reduction
is significant, the main contribution comes from the electrons that are
injected above the potential barrier height by the value about $k_BT$.
Therefore, the integral noise-reduction effect is a consequence of
the suppression of the contributions from the electron energies
in the vicinity of $\varepsilon^*$.

The obtained exact solutions allows us to investigate in great detail
the correlations between different groups of carriers.
While the injected carriers are uncorrelated, those in the volume of the
conductor are strongly correlated, as follows from the derived formulas
for the fluctuation of the distribution function [see general expressions
(\ref{ddfinj}) and (\ref{ddfind}) in the Appendix].
Those correlations may be observed experimentally by making use of
a combination of two already realized techniques: a hot-electron
spectrometer \cite{hayes85,heiblum85} and shot-noise measurements.
\cite{reznikov95,kumar96,schoel97}
The electron spectrometer, placed behind the receiving semitransparent
contact, acts as an analyzer of electron distribution over the energy.
\cite{hayes85,heiblum85}
In this way spectroscopic information, that is, the average partial currents
$\bar{I}(\tilde{\varepsilon})$ and their fluctuations 
$\delta I(\tilde{\varepsilon})$, may be measured for different energies 
$\tilde{\varepsilon}$ of electrons collected at the contact.
This is similar to the energy-resolved noise measurements realized in
Ref.\ \onlinecite{yao98}.
The partial current of the transmitted electrons at the receiving (right)
contact is given by $\bar{I}(\tilde{\varepsilon})=
I_c e^{-\tilde{\varepsilon}-V_m}\theta(\tilde{\varepsilon})$,
where the threshold energy $\tilde{\varepsilon}$=0 corresponds to
the arriving electrons that have a zero longitudinal kinetic energy
at the potential minimum.
To find the fluctuation $\delta I(\tilde{\varepsilon})$,
we consider the fluctuation of the distribution function
$\delta f(x,w)$ at $x=\lambda$.
Since $\delta\psi(\lambda)$=0, the terms with $\delta\psi(x)$ vanish.
Thus, for the transmitted over the barrier electrons which contribute
to the current, from Eqs.\ (\ref{dfinjxt}) and (\ref{dfindxt}) one obtains

\begin{eqnarray}
\delta f_{L,t}(\lambda,w) &=&
\delta f_L(\lambda,w) \, \theta(w-w_R) \nonumber\\
&&- \frac{1}{\sqrt{\pi}} e^{-w^2 + V}\,\delta V_m\frac{1}{2w}\delta(w-w_R).
\end{eqnarray}
Since only the positive velocities are considered, one can change the
velocity variable to the energy by $\tilde{\varepsilon}=w^2-w_R^2$, and obtain

\begin{equation}
\delta f_{L,t}(\tilde{\varepsilon}) =
\delta f_L(\tilde{\varepsilon}+V_m)\, \theta(\tilde{\varepsilon})
- \frac{1}{\sqrt{\pi}} e^{-\tilde{\varepsilon} - V_m}\,\delta V_m
\delta(\tilde{\varepsilon}).
\end{equation}
By using the relation (\ref{di}) between the fluctuation of the contact 
distribution function and that of the contact injection current, we obtain

\begin{equation}
\delta I(\tilde{\varepsilon}) =
\delta I_L(\tilde{\varepsilon}+V_m)\, \theta(\tilde{\varepsilon})
- I_c\,e^{-V_m}\,\delta V_m\,\delta(\tilde{\varepsilon}).
\end{equation}
Thus, the correlator for the current fluctuations becomes

\begin{eqnarray}
\langle\delta I(\tilde{\varepsilon})\,\delta I(\tilde{\varepsilon}')\rangle
|_{x=\lambda}
&=&\langle\delta I_L(\tilde{\varepsilon}+V_m)\,
\delta I_L(\tilde{\varepsilon}'+V_m) \rangle \nonumber\\
&&- I_c\,e^{-V_m}\,\delta(\tilde{\varepsilon}')\,
\langle\delta I_L(\tilde{\varepsilon}+V_m)\,\delta V_m\rangle \nonumber\\
&&- I_c\,e^{-V_m}\,\delta(\tilde{\varepsilon})\,
\langle\delta I_L(\tilde{\varepsilon}'+V_m)\,\delta V_m\rangle \nonumber\\
&&+ I_c^2\,e^{-2V_m}\,\delta(\tilde{\varepsilon})\,
\delta(\tilde{\varepsilon}')\,\langle\delta V_m^2\rangle,
\end{eqnarray}
where the average is taken over the injected fluctuations.
It is clear that for $\tilde{\varepsilon}, \tilde{\varepsilon}'>0$
the carriers remain uncorrelated since only the first term does not vanish.
It is $\propto\delta(\tilde{\varepsilon}-\tilde{\varepsilon}')$ due to
the imposed injection conditions that should lead to the full shot noise.
In such a case, an interesting question arises:
What is the reason for the noise reduction obtained for the total
(integrated over the energies) current fluctuations?
The answer is found looking at the electrons with energies close to the
threshold energy $\tilde{\varepsilon}$=0 (``tangent'' electrons).
All other electrons are anticorrelated with that group.
This means that if there is a positive fluctuation of overbarrier electrons,
there should be a negative one for the ``tangent'' electrons and vice versa.
This anticorrelation explains the overall noise reduction.
The tangent electrons can be thought as overcorrelated.
The dispersion $\langle\delta I^2(\tilde{\varepsilon})\rangle$
has a sharp peak at $\tilde{\varepsilon}$=0 and then decreases with energy
at $\tilde{\varepsilon} > 0$.
This peak is divergent ($\delta$-shaped) in our collisionless theory.
A small probability of scattering will lead to its broadening
and finite magnitude.
Therefore, by measuring the dispersion of the partial current fluctuations
and/or their cross-correlations, one may observe a
sharp peak and an anticorrelation of electrons,
thus making the Coulomb correlations effect visible.

\section{Summary}

In conclusion, we have presented a self-consistent theory of electron 
transport and noise in a ballistic two-terminal conductor under 
the conditions of nondegenerate electron gas. Our description is valid 
for ballistic electrons in solids as well as in vacuum.
By solving analytically the kinetic equation coupled self-consistently
with a Poisson equation, we have derived the electron distribution function
and its fluctuation at arbitrary section $x$ of the conductor.
This allowed us to obtain the steady-state spatial distributions of
the transport characteristics, the $I$-$V$ curves, and the noise
characteristics.
While the time-averaged quantities are not affected by the Coulomb
correlations, the noise characteristics are demonstrated to be drastically
modified when those correlations are taken into account.
Our results are in excellent agreement with the preceding Monte Carlo
simulations. \cite{gonzalez97,gonzalez98b}

The obtained formulas have been analyzed in a wide range of biases
and compared with the correspondent theory for the vacuum diode.
In particular, we have demonstrated that the known formulas for vacuum
electronics, such as the Child 3/2-power law for $I$-$V$ characteristics or
North's asymptotic formula for the noise may not be applied for
the semiconductor diode at biases that are relevant for the ballistic
transport regime.
Instead, one should use the more general formulas described in the present
paper from which follows
(i) the linear or sublinear $I$-$V$ dependences even under a strong limitation
of transport by a space charge;
(ii) a noise level significantly below the level obtained from the North's
formula;
(iii) the sharp recovering of the full shot-noise level at a certain
critical voltage;
(iv) all the transport and noise characteristics determined by two
dimensionless parameters: the screening parameter $\lambda=d/L_D^0$ and
the bias $qU/k_BT$.

Based on the derived formula for the current-noise spectral density
one may distinguish the relative contributions to the noise from different
groups of carriers.
We have analyzed the contributions coming from the transmitted and reflected
groups of carriers, as well as the partial contributions to the noise from
electrons of different injection energies.
Finally, it should be noted that the analytical approach that we have
presented in the paper may be extended and applied to various systems,
e.g., for different kind of statistics of injecting electrons, \cite{unpub}
and other types of the contacts.
On the other hand, the shot-noise suppression effect, which we treat
analytically, may lead to important applications for low-noise small-size
semiconductor devices, generators of sub-Poissonian light sources,
\cite{saleh92} etc.
Our work then offers new perspectives on the study of Coulomb interactions
and noise in small-size ballistic devices, such as ballistic transistors, 
point contacts, etc.

\acknowledgements

We are grateful to T.\ Gonz\'alez and L.\ Reggiani for fruitful collaboration
on the Monte Carlo investigation of the present problem.
This work has been partially supported by
the Direcci\'on General de Ense\~nanza Superior, Generalitat de Catalunya,
Spain, and the NATO linkage grant HTECH.LG 974610.

\appendix

\section{Fluctuations of the distribution function and electron density}

In a similar way to the subdivision of thestationary distribution function 
(\ref{dfsum}) into the components corresponding to different groups
of electrons classified in Sec.\ \ref{stat-df},
the fluctuation $\delta f(x,w)$ may be expressed as

\begin{equation} \label{dffsum}
\delta f = \delta f_{L,t} + \delta f_{L,r} + \delta f_{R,t} + \delta f_{R,r}.
\end{equation}
The boundary conditions for these functions are obtained by perturbing
the steady-state boundary conditions (\ref{bcdf}) and using
$\partial w_k/\partial\psi_m = -(2w_k)^{-1}$. One gets

\begin{mathletters} \label{bcdff} \begin{eqnarray}
\delta f_{L,t}(0,w_c) &=& \delta f_L(w_c) \, \theta(w_c - w_L)  \nonumber\\
&&+ \frac{1}{2w_L} f_L(w_c) \, \delta(w_c - w_L)\,\delta\psi_m,
\label{bcdffa}\\
\delta f_{L,r}(0,w_c) &=& \delta f_L(w_c) \,\theta(w_L^2 - w_c^2)
\nonumber\\ &&- f_L(w_c) \, \delta(w_L^2 - w_c^2)\, \delta\psi_m,
\label{bcdffb}\\
\delta f_{R,t}(\lambda,w_c) &=& \delta f_R(w_c) \, \theta(-w_c - w_R)
\nonumber\\ &&- \frac{1}{2w_R} f_R(w_c) \,
\delta(-w_c - w_R)\,\delta\psi_m, \label{bcdffc} \\
\delta f_{R,r}(\lambda,w_c) &=& \delta f_R(w_c) \,\theta(w_R^2 - w_c^2)
\nonumber\\ &&- f_R(w_c) \, \delta(w_R^2 - w_c^2)\, \delta\psi_m.
\label{bcdffd}
\end{eqnarray} \end{mathletters}
where the additional terms proportional to $\delta\psi_m$ describe
the changes in the distribution functions due to the potential barrier
variation.

Now we have to solve the perturbed kinetic equation (\ref{kinf}),
which may be rewritten as

\begin{equation} \label{kinf2}
\left( w \,\frac{\partial}{\partial x} + \frac{1}{2} \frac{d\psi}{dx}\,
\frac{\partial}{\partial w} \right) \delta f(x,w)
= - \frac{1}{2} \frac{\partial f}{\partial w}\,\frac{d \delta \psi}{dx},
\end{equation}
where the rhs is supposed to be a given function
(for this step of calculations).
A general solution of this nonhomogeneous partial differential equation
is a sum of a solution of the homogeneous problem
and a particular solution of the nonhomogeneous problem. Explicitly,

\begin{equation} \label{dftot}
\delta f_{k,j} = \delta f_{k,j}^{hom} + \delta f_{k,j}^{nhom},
\quad k=L,R, \quad j=t,r.
\end{equation}

The solution for the homogeneous problem is determined
by the boundary conditions (\ref{bcdff}).
By making use of the energy-conservation law (\ref{encons}), we make a
replacement

\begin{equation} \label{wc}
w_c = {\rm sgn}(w)\,\sqrt{w^2-\psi(x)+\psi_k}
\end{equation}
and obtain different contributions to $\delta f_{k,j}^{hom}$ in the form

\begin{mathletters} \label{ddfhom} \begin{eqnarray}
\delta f_{k,t}^{hom}(x,w) &=& \delta f_k(x,w) \, 
\theta\biglb(\pm w-w_*(x)\bigrb) \label{ddfhoma}\nonumber\\
&& \pm \frac{1}{2w} f_k(x,w)\, \delta\biglb(\pm w-w_*(x)\bigrb)\,
\delta\psi_m,\\
\delta f_{k,r}^{hom}(x,w) &=& \delta f_k(x,w) \,
\theta\biglb(w_*^2(x)-w^2\bigrb) \nonumber\\ 
&&- f_k(x,w)\, \delta\biglb(w_*^2(x)-w^2\bigrb)\,\delta\psi_m, \label{ddfhomb}
\end{eqnarray} \end{mathletters}
where $\delta f_{L,r}^{hom}$ and $\delta f_{R,r}^{hom}$ are defined
in the regions $0<x<x_m$ and $x_m<x< \lambda$, respectively.
The upper sign applies for $\delta f_{L,t}^{hom}$ and
the lower sign applies for $\delta f_{R,t}^{hom}$, both terms valid
in the whole range $0<x< \lambda$.
The critical velocity $w_*(x)$ is given by Eq.\ (\ref{wcrit}).

The solution of the nonhomogeneous problem can easily be found through 
the steady-state distribution function 
$f(\varepsilon_t)=f\biglb(w^2-\psi(x)\bigrb)$
in terms of the total energy $\varepsilon_t$ or, equivalently,
in terms of the injection velocity $w_c$,

\begin{equation} \label{ddfnhom1}
\delta f_{k,j}^{nhom}=
-\frac{\partial f_{k,j}}{\partial\varepsilon_t}\,
\delta\psi = -\frac{1}{2w_c}\,
\frac{\partial f_{k,j}}{\partial w_c}\, \delta\psi.
\end{equation}
Differentiating Eqs.\ (\ref{bcdf}), we find

\begin{mathletters} \label{ddfnhom} \begin{eqnarray}
\delta f_{L,t}^{nhom} &=& f_L(w_c) \,\delta\psi(x) \nonumber\\
&& \times \left[\theta(w_c-w_L)
- \frac{1}{2w_c} \delta(w_c-w_L)\right], \\
\delta f_{R,t}^{nhom} &=& f_R(w_c) \,\delta\psi(x) \nonumber\\
&& \times \left[\theta(-w_c-w_R)
+ \frac{1}{2w_c} \delta(-w_c-w_R)\right], \\
\delta f_{k,r}^{nhom} &=& f_k(w_c) \,\delta\psi(x) \nonumber\\
&& \times \left[\theta(w_k^2-w_c^2)+\delta(w_k^2-w_c^2)\right].
\end{eqnarray} \end{mathletters}
In these equations the substitution (\ref{wc}) is assumed,
so that the fluctuations are finally the functions of ($x,w$).
Notice that the components for the reflected groups of carriers are defined
in the regions: $\delta f_{L,r}^{nhom}$ for $0<x<x_m$ and
$\delta f_{R,r}^{nhom}$ for $x_m<x< \lambda$, while those for the
transmitted groups of carriers are given in the whole range $0<x< \lambda$.

According to the electrostatic boundary conditions (\ref{bcpoisf})
the fluctuations of the potential at the contacts are equal to zero, which
leads to vanishing contributions (\ref{ddfnhom}) at the
contacts $\delta f^{nhom}(0,w)=\delta f^{nhom}(\lambda,w)=0$.
The contributions (\ref{ddfhom}) satisfy the boundary conditions (\ref{bcdff}).
Thus, the distribution function in the form (\ref{dftot}) with eight
contributions (\ref{ddfhom}) and (\ref{ddfnhom}) is the solution of the problem
for a given electrostatic potential $\psi(x)+\delta\psi(x)$.

For convenience of further consideration, we present $\delta f$ as a sum of
the ``injected'' and ``induced'' contributions

\begin{equation} \label{dftot2}
\delta f_{k,j} = \delta f_{k,j}^{inj} + \delta f_{k,j}^{ind},
\quad k=L,R, \quad j=t,r.
\end{equation}
In terms of the contact velocities $w_c$ (presented in such a form
these equations will be frequently used throughout of the paper),
those contributions are given by

\begin{mathletters} \label{ddfinjc} \begin{eqnarray}
\delta f_{k,t}^{inj}(w_c) &=& \delta f_k(w_c) \, 
\theta\biglb(\pm w_c-w_k\bigrb), \\
\delta f_{k,r}^{inj}(w_c) &=& \delta f_k(w_c) \,
\theta\biglb(w_k^2-w_c^2\bigrb),
\end{eqnarray} \end{mathletters}
and

\begin{mathletters} \label{ddfindc} \begin{eqnarray}
\delta f_{k,t}^{ind}(x,w_c) &=& f_k(w_c) \left\{\theta(\pm
w_c-w_k)\delta\psi(x)\right.
\nonumber\\&& \left. \hspace{-6mm} \mp \frac{1}{2w_c}
\delta(\pm w_c-w_k)[\delta\psi(x)-\delta\psi_m]\right\}, \\ \delta
f_{k,r}^{ind}(x,w_c) &=& f_k(w_c)
\left\{\theta(w_k^2-w_c^2)\delta\psi(x)\right.
\nonumber\\&& \left.
+\delta(w_k^2-w_c^2)[\delta\psi(x)-\delta\psi_m]\right\}.
\end{eqnarray} \end{mathletters}
where the substitution (\ref{wc}) is assumed.
The same terms as functions of $(x,w)$ are determined by the formulas

\begin{mathletters} \label{ddfinj} \begin{eqnarray}
\delta f_{k,t}^{inj}(x,w) &=& \delta f_k(x,w) \, 
\theta\biglb(\pm w-w_*(x)\bigrb), \label{dfinjxt} \\
\delta f_{k,r}^{inj}(x,w) &=& \delta f_k(x,w) \,
\theta\biglb(w_*^2(x)-w^2\bigrb),
\end{eqnarray} \end{mathletters}
and

\begin{mathletters} \label{ddfind} \begin{eqnarray}
\delta f_{k,t}^{ind}(x,w) &=& \frac{1}{\sqrt{\pi}} e^{-w^2 + \psi(x) -\psi_k}
\left\{\theta\biglb(\pm w-w_*(x)\bigrb)\delta\psi(x)\right. \nonumber\\
&& \left. \hspace{-6mm} \mp \frac{1}{2w} \delta\biglb(\pm w-w_*(x)\bigrb)
[\delta\psi(x)-\delta\psi_m]\right\}, \label{dfindxt}\\
\delta f_{k,r}^{ind}(x,w) &=& \frac{1}{\sqrt{\pi}} e^{-w^2 + \psi(x) -\psi_k}
\left\{\theta\biglb(w_*^2(x)-w^2\bigrb)\delta\psi(x)\right.  \nonumber\\&& 
\left. +\delta\biglb(w_*^2(x)-w^2\bigrb)[\delta\psi(x)-\delta\psi_m]\right\}.
\end{eqnarray} \end{mathletters}
Apparently, $\delta f^{inj}$ has a meaning of the distribution function of
randomly {\em injected} electrons, while $\delta f^{ind}$ describes the change
in the steady-state distribution {\em induced} by injected electrons.

The obtained fluctuations of the distribution function
allows one to compute each contribution to the fluctuations of the electron
density $\delta n(x)$ by integrating over velocities.
Changing the integration over $w$ to that over the contact injection velocities
$w_c$, we find

\begin{eqnarray}
\delta n_{k,j}(x) &=& \int_{-\infty}^{\infty} \delta f_{k,j}(x,w) dw
\nonumber\\
&=& \int_{-\infty}^{\infty} \frac{\delta f_{k,j}(w_c) w_c d w_c}
{{\rm sgn}(w_c)\,\sqrt{w_c^2+\psi(x)-\psi_k}}.
\end{eqnarray}
Thus, by using Eqs.\ (\ref{ddfinjc}) and (\ref{ddfindc}), one obtains
for the injected density fluctuations

\begin{mathletters} \label{dninj} \begin{eqnarray}
\delta n_{L,t}^{inj}(x) &=& \int_{w_L}^{\infty}
\frac{\delta f_L(w_c) w_c \,d w_c}{\sqrt{w_c^2+\psi(x)-\psi_L}}, \\
\delta n_{R,t}^{inj}(x) &=& -\int_{-\infty}^{-w_R}
\frac{\delta f_R(w_c) w_c \,d w_c}{\sqrt{w_c^2+\psi(x)-\psi_R}}, \\
\delta n_{k,r}^{inj}(x) &=& 2\int_{\sqrt{\psi_k-\psi(x)}}^{w_k}
\frac{\delta f_k(w_c) w_c\, d w_c}{\sqrt{w_c^2+\psi(x)-\psi_k}},
\end{eqnarray} \end{mathletters}
and for the induced fluctuations

\begin{mathletters} \label{dnind} \begin{eqnarray}
\delta n_{k,t}^{ind}(x) &=& n_{k,t}\, \delta\psi(x)
-\frac{e^{-\psi_k-\psi_m}}{2\sqrt{\pi}w_*(x)} [\delta\psi(x)-\delta\psi_m], \\
\delta n_{k,r}^{ind}(x) &=& n_{k,r}\, \delta\psi(x)
+\frac{e^{-\psi_k-\psi_m}}{\sqrt{\pi}w_*(x)} [\delta\psi(x)-\delta\psi_m],
\end{eqnarray} \end{mathletters}
Here, the contributions (\ref{dninj}) can be interpreted as
the electron-density fluctuations at a slice $x$ caused by the stochastic
injection from the contacts to the base.
The contributions (\ref{dnind}) are related to a variation of the stationary
electron density due to a local variation of the potential and its minimal
value (a self-consistent response).
As before, the terms $\delta n_{L,r}$ and $\delta n_{R,r}$ are defined
on the intervals $0<x<x_m$ and $x_m<x< \lambda$, respectively,
while the terms $\delta n_{k,t}$ are defined on the whole range $0<x< \lambda$.

\end{multicols}

\begin{references}
\bibitem{remark1}
The closest analog from vacuum electronics is a double-cathode valve
\cite{knol50} containing two hot cathodes under the same temperature opposite
each other and operating under a space-charge-limited transport regime;
see also Ref.\ \onlinecite{apl99}.
\bibitem{remark2}
It follows from the fact that the Coulomb potential is a power function
having no characteristic scale. The distance between the contacts may be
chosen as the length scale for estimates.
\bibitem{remark3}
An electron with $\varepsilon$=$V_m$ comes to rest in the virtual cathode,
producing an infinitely large perturbation of the current.
However, this effect is unphysical, since that point corresponds
to an unstable equilibrium for the electron.
The physically meaningful quantity is the current noise obtained by the
integration over all injection energies, and that integral is well defined
because of the logarithmic type of the singularity.
\end{references}
\end{document}